\documentclass[twocolumn,showpacs,preprintnumbers,amsmath,amssymb]{revtex4}

\usepackage{graphicx}
\usepackage{graphicx,epstopdf}
\usepackage{dcolumn}
\usepackage{bm}
\usepackage{mathrsfs} 

\newcommand{\ben}{\begin{enumerate}}
	\newcommand{\een}{\end{enumerate}}

\usepackage{color}

\begin{document}

\title{Survivability of a metapopulation under local extinctions}
\author{Srilena Kundu$^1$}
 \author{Soumen Majhi$^1$}
 \author{Sourav Kumar Sasmal$^2$}
\author{Dibakar Ghosh$^1$}
\email{diba.ghosh@gmail.com}
\author{Biswambhar Rakshit$^3$}
\affiliation{$^1$Physics and Applied Mathematics Unit, Indian Statistical Institute, Kolkata-700108, India\\
	$^2$Agricultural and Ecological Research Unit, Indian Statistical Institute, Kolkata-700108, India\\
	$^3$ Department of Mathematics, Amrita School of Engineering-Coimbatore Amrita Vishwa Vidyapeetham, India}

\date{\today}

\begin{abstract} 
A metapopulation structure in landscape ecology comprises a group of interacting spatially separated subpopulations or patches of the same species that may experience several local extinctions. This makes the investigation of survivability (in the form of global oscillation) of a metapopulation on top of diverse dispersal topologies extremely crucial. However, among various dispersal topologies in ecological networks, which one can provide higher metapopulation survivability under local extinction is still not well explored. In this article, we scrutinize the robustness of an ecological network consisting of prey-predator patches having Holling type I functional response, against progressively extinct population patches. We present a comprehensive study on this while considering global, small-world and scale-free dispersal of the subpopulations. Furthermore, we extend our work in enhancing survivability in the form of sustained global oscillation by introducing asymmetries in the dispersal rates of the considered species. Our findings affirm that the asynchrony among the patches plays an important role in the survivability of a metapopulation. In order to demonstrate the model independence of the observed phenomenon, we perform a similar analysis for patches exhibiting Holling type II functional response. On the grounds of the obtained results, our work is expected to provide a better perception of the influence of dispersal arrangements on the global survivability of ecological networks. 
	        
\end{abstract}


\maketitle


\section{Introduction}
\par  Collective behavior in a large ensemble of coupled oscillators has been given great importance in recent times due to its applicability in modeling various self-organized complex systems \cite{Kurths-book,Strogatz}. The emergent dynamics of such complex systems depends both on coupling topology as well as individual elements. In recent times, much attention has been given to understanding the dynamical robustness in a mixed population \cite{Tanaka2014,Tanaka2015,Tanaka2015a,Tanaka2017,daido2004,tanaka2012sr,liu2016,tanaka2010p,daido2007}. Such study of dynamical robustness is very much relevant in various biological as well as ecological systems.  It is quite natural that degradation may happen due to aging in some  living systems. If this degradation reaches a certain critical level, the proper functioning of a large system may hamper and face severe disruption. Therefore, it is necessary to examine up to which level such a large system can survive against such deteriorations.

In ecology, prey-predator systems have attracted great interest due to a variety of complex biological processes. In spatial ecology, the movement of spatially separated populations of the same species is described  by metapopulation dynamics \cite{levin1970}. Research on the long-term dynamics of spatially structured populations has revealed that population densities of a given species tend to fluctuate in synchrony over vast geographical areas \cite{Ranta95}. In metapopulation dynamics, a patch is generally modeled as a system of differential equations which exhibits an oscillating solution \cite{Goldwyn}. Therefore it is quite natural to suggest that spatially structured biological populations can be viewed as a network of coupled oscillators. In this representation, nodes represent suitable habitat patches and links between such nodes indicate proper pathways. It may happen that because of  climate variations,  environmental heterogeneity, and many other reasons, habitats in a particular patch may go for extinction.

Network science has witnessed many developments in the last decade due to a variety of topological structures found in many real-world networks \cite{Watts-Strogatz}. Like physicists, ecologists are also placing more importance on the study of networks of coupled oscillators. The rationale behind studying complex networks lies in the fact that the network topology highly affects its dynamics and some recent papers  have explicitly investigated the dynamical consequences of network patterns \cite{Ranta2008,Gilarranz2012,sinha201}.

 Most of the theoretical studies on patchy environments \cite{levin1970, levin1974, hastings1993} have concentrated on negative density dependent local dynamics (i.e., population fitness will be maximum at low density) and density independent dispersal. But in reality, the reverse holds true: individuals of many species cooperate, when the population density is low. Individuals use cooperative strategies to hunt, they forage together, they fight together to survive under unfavorable conditions, or they seek sexual reproduction at the same moment. It may be that they will benefit from more resources but in many cases, they will also suffer from a lack of conspecifics, at low density. If this is stronger than the benefits, then individuals may be less likely to survive.  Experimental observation on the flour beetle, \textit{Tribolium confusum}, shows that per capita growth rate (pgr) was highest at their intermediate densities (unlike logistic growth). Thus, in this paper, we consider a Rosenzweig-MacArthur prey-predator model with positive density dependent growth of prey, i.e., subject to the Allee effect, which refers to the positive correlation between population size/density and its pgr at low population density \cite{Allee,Franck2008}.


The Allee effects are mostly classified into two categories: strong and weak Allee effects. In strong Allee effects, there is a threshold density (Allee threshold), below which the pgr becomes negative and extinction  becomes certain.  On the other hand, the Allee effect without this threshold density is known as the weak Allee effect and in this case the population always exhibits a positive pgr.	Thus the strong Allee growth function increases the chance of extinction of prey and thus of predators in some patches. Here we focus on ecological robustness of networks of coupled patches, which is defined as the ability of the whole network to support metapopulation persistence when  in a fraction of habitat patches both prey and predators become extinct. But still, which dispersal topology supports more metapopulation persistence is not well investigated, to the best of our knowledge.  For this we consider three different network structures, namely two homogeneous (global and small-world network) and one heterogeneous network (scale free). We comparatively study the ecological robustness using these three networks. First, we consider a global (all-to-all) network where species dispersal is happening from each patch to every other patch.  Next we consider a small-world network which is a homogeneous network with random interaction among the patches. Regular networks consist of interactions among neighboring oscillators, which are then rewired with probability $q$ between randomly selected nodes in the network. If $q=0$, it represents a completely regular network, while $q=1$ gives a random network.  The small-world network is somewhere $0<q<1$, $q$ being sufficiently small maintaining a high clustering coefficient and low average path length of the network. Both the global network as well as small-world networks are highly homogeneous as far as the degree distributions of the nodes are concerned. Finally we consider a scale-free network which is extremely heterogeneous where a majority of nodes have one or a few links, but a few nodes are extremely well connected. We consider all these networks one-by-one and investigate how the ecological robustness varies depending on the network dispersal structure. For all types of dispersal configurations, we observe that revival of species takes place in inactive patches due to dispersal until the number of inactive patches reaches a critical value. We also notice that  in the  case of small-world networks, the chances of metapopulation persistence are much higher compared to global as well as scale-free networks.

\par  The rest of the paper is organized as follows: The general mathematical description of the network is provided in Sec. II. In Sec. III, we first briefly describe the Rosenzweig-MacArthur prey-predator model with a Holling type I functional response and discuss the ecological significance of the parameters. Then, we present our results on global, small-world, and scale-free ecological networks. In Sec. IV, we validate the observed results through the analysis on the metapopulation of patches having a Holling type II functional response. We summarize our results with a discussion in Sec. V.

\section{Network of Coupled Patches}

We consider a network of $N$ patches with dispersal in terms of diffusive coupling. This means that populations, which are spatially isolated, are linked by dispersal using diffusive coupling. Thus the prey-predator model of $N$ patches is represented by the following equation,
\begin{equation}\label{eq:2}
\dot{\mathbf{X}_i} = \mathbf{F(X}_i) + \mathbf{M} \sum_{j=1}^{N}A_{ij}(\mathbf{X}_j-\mathbf{X}_i),
\end{equation}
where $\mathbf{X}_i = (x_i, y_i)^T $ denotes the state vector, $\mathbf{F(X}_i) = (f(x_i,y_i), g(x_i,y_i))^T$ describes the inherent dynamics of the $i$th patch for $i=1, 2, ..., N$, and $T$ denotes the transpose of a matrix. The second term is the diffusive coupling that denotes interaction among the species of different patches.  Here $\mathbf{M}=\Big(\frac{\alpha m}{deg(i)}, \frac{\beta m}{deg(i)}\Big)^T$ denotes the dispersal matrix where $m$ is the dispersal rate between the patches; $\alpha$, $\beta$ are the asymmetry constants induced in the dispersal; and $deg(i)$ is the number of patches (i.e., degree) connected with the $i$th patch.   $A_{ij}$ represents the adjacency matrix where $A_{ij}=1$ if dispersal is happening between the $i$th and $j$th patches; otherwise $A_{ij}=0$.  In our network model, we assume that the dispersal is instantaneous (i.e., no time delay during dispersal) and there is no birth or death during dispersal.  Here we consider that in a fraction $p$ of patches, species are extinct when there is no dispersal among the patches. That means that out of $N$ patches, in a $pN$ number of patches,  species are in the extinct state (we call them inactive or unfit patches), whereas in a $(1-p)N$ number of patches they are in the oscillatory state (active or fit patches).
Whenever a fraction $p$ of the patches become extinct due to some degradations, the amplitude of the global oscillation decreases. As $p$ increases beyond a certain threshold $p_c$ (say), the oscillatory behavior of the metapopulation suddenly vanishes to a global extinction. This critical $p_c$ value is used as a measure of metapopulation survivability.
In order to study the metapopulation survivability, we consider three different dispersal topologies, namely, global, small-world, and scale-free networks, and examine the ecological robustness of the whole network as we increase the fraction $p$ of extinct species patches from 0 to 1. For this purpose, we define an order parameter to measure the dynamical activities in the network. The order parameter $R$ is defined as
\begin{equation}
R = \frac{1}{2}(R_x+R_y),
\end{equation}
where $R_x = \frac{1}{N}\sum_{i=1}^{N}(\langle x_{i,max}\rangle_t-\langle x_{i,min}\rangle_t)$,  $R_y=\frac{1}{N}\sum_{i=1}^{N}(\langle y_{i,max}\rangle_t-\langle y_{i,min}\rangle_t)$, and $\langle...\rangle$ defines a  long time average.  This order parameter $R=0$ signifies the existence of stable steady states which may be trivial (zero equilibrium) or nontrivial.  To distinguish between trivial (extinction state) and nontrivial steady states, we introduce $\Delta = \Theta(\bf{X_i} - \delta)$ where $\delta$ is a  predefined threshold and $\Theta(x)$ is the Heaviside step function.  We choose $\delta=10^{-6}$ so that for the extinction state, $\Delta$ gives a value zero and for the nontrivial steady state, $\Delta=1$.   This $R$ identifies the average amplitude of the networked system. Nonzero values of the order parameter $R$ imply  persistence  of a metapopulation in the whole network while $R=0$ signifies extinction of species (as the global oscillation vanishes) from all the patches in the network with $\Delta=0$.
\par In the following, we explore the survivability of a  metapopulation while estimating the critical inactivation ratio $p_c$ for global, small-world and scale-free dispersal networks. Our motivation is to explain how this critical threshold $p_c$ varies with dispersal rate $m$ over these dispersal networks.

\section{Rosenzweig-MacArthur prey-predator model with Holling type I functional response}
We consider the Rosenzweig-MacArthur prey-predator model in the presence of the strong Allee effect. The mathematical form of single-patch dynamics \cite{Sasmal2017} is given by 
\medskip
\begin{equation}\label{eq:1}
\begin{split}
\frac{dx}{dt} & = f(x, y)=\frac{1}{\epsilon}\Big[x(1-x)(x-\theta)-xy\Big]\\
\frac{dy}{dt} & = g(x, y)=xy-dy
\end{split}
\end{equation}
where $x$ and $y$ are the normalized prey and predator population density, respectively. Here $\epsilon\in(0,1]$ denotes the time scale difference between prey and predator populations, $\theta\in(0,1)$ denotes the Allee threshold that corresponds to a critical population size or density below which the population will be settled for extinction without any further abatement, and $d$ is the natural mortality rate of the predator population.


\begin{figure}[ht]
	\begin{center}
		\includegraphics[scale=0.5]{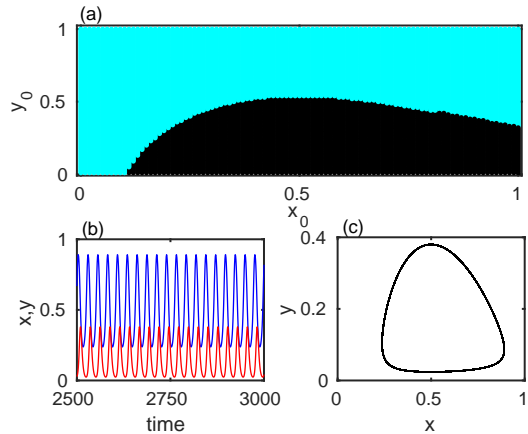}
		\caption{ Dynamics of the single Rosenzweig-MacArthur model Eq. (1). (a) basin of attraction for $d = 0.5, \theta=0.1$, and $\epsilon= 1$. Color cyan represents initial conditions corresponding to extinction (zero equilibrium) state and initial conditions from the black region lead to  oscillatory state. (b) Time series of prey (blue line)  and predator (red line) populations and (c) corresponding phase space shows limit cycle attractor.}
		\label{fig:fig1}
	\end{center}
\end{figure}

\par  For the single-patch model \eqref{eq:1}, the interior equilibrium $(d,(1-d)(d-\theta))$, which exists if $\theta<d<1$, is stable if $d>\frac{1+\theta}{2}$ and there is a supercritical Hopf bifurcation at $d=\frac{1+\theta}{2}$. For $\theta<d\leq\frac{1+\theta}{2}$,  coexistence of oscillation and the stable extinction state $(0,0)$ in both populations occurs depending on the initial population density. For $d\leq\theta$, the system converges to $(0,0)$ for any initial conditions in $R^2_+$. This phenomenon is known as overexploitation or predator-driven extinction. In the presence of strong Allee effects, due to large predator invasion, predator growth is fast enough to drive the prey below the Allee threshold and leads to the extinction of both  populations \cite{van2007h}.  Thus a higher value of $d$ shows the stable coexistence of all the populations, and as we decrease the value of $d$, the system behavior changes from stable coexistence to extinction (via the  oscillatory state) for all the populations. For these reasons we considered the parameter $d$ as a measure of the patch quality. In our present study, an active patch means both populations are in stable limit cycle oscillations. The basin of attraction for $d = 0.5, \theta=0.1$, and $\epsilon= 1$ in the single-patch model \eqref{eq:1} is shown in Fig.~\ref{fig:fig1}(a). Initial conditions leading to extinction state $(0, 0)$ and responsible for the oscillatory state are plotted in cyan and black, respectively. The stable time series of prey $x(t)$ and predator $y(t)$ and the corresponding phase plane plot depicting the fluctuation in prey and predator populations for an  isolated Rosenzweig MacArthur model are shown in Figs.~\ref{fig:fig1}(b) and (c), respectively.

\subsubsection{Global Dispersal}
First we consider global (all-to-all) dispersal  and investigate the survivability of a metapopulation when in a fraction of patches species become extinct. This means that in the case of a global network, dispersal takes place uniformly in all the selected patches; i.e., prey and predators are moving from each patch to all other patches equally. Without loss of generality, we can set the group of extinct patches for ${j=1, 2,...,Np}$ and the rest of the patches as active for ${j=Np+1,...,N}$.  For our numerical simulations \cite{scheme}, we set $N=200$ as the total number of patches in the network. Throughout the paper, we set the natural mortality rate $d=0.2$ for inactive patches and $d=0.5$ for active patches. We take random initial conditions for active patches in the interval $(0.2, 1.0)$ and $(0.0, 0.3)$ for prey and predator populations respectively.  For inactive patches at $d=0.2$, random initial conditions from the interval $[0, 1]$ are chosen for both prey and predator populations.  With these system parameters and initial conditions, an isolated patch exhibits either a stable steady state at the origin (inactive patch) or a stable limit cycle around the coexistence equilibrium (active patch).
\begin{figure}[ht]
	\begin{center}
		\includegraphics[scale=0.6]{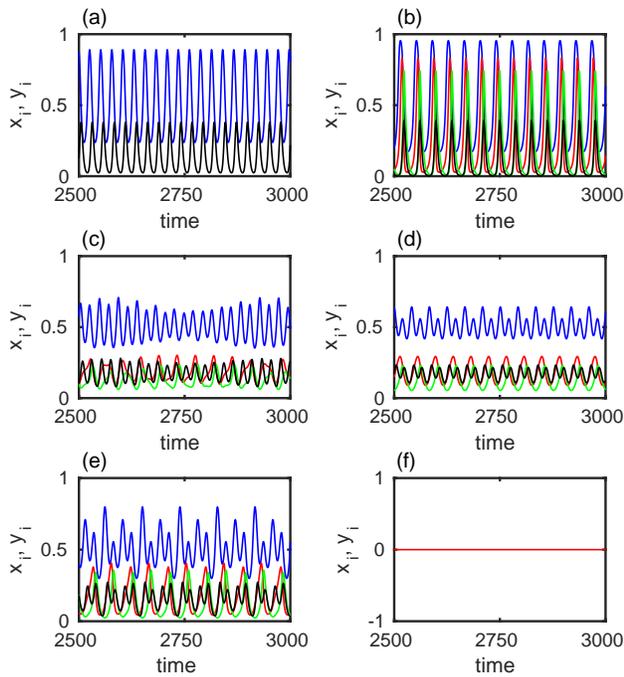}
		\caption{Time series for different values of the inactivation ratio: (a) periodic $p=0.0$, (b) periodic with highest amplitude $p=0.15$, (c) chaotic $p=0.50$, (d) period 2 (active) and period 1 (inactive) $p=0.6$, (e) period 4 (active) and period 2 (inactive) $p=0.7$, and (f) extinction state $p=0.725$.  Blue and black lines represent the prey and predator population in active patches and for inactive patches they are shown by red and green lines, respectively.  Here $N=200$, $m=0.07$, and $\alpha=\beta=1$ are fixed.
			\label{fig:fig2}}
	\end{center}
\end{figure}	  
\par We fix the value of the dispersal rate for the global network at $m=0.07$ (for the identical dispersal case $\alpha=\beta=1.0$) and vary the inactivation parameter $p$.  	
In Fig.~\ref{fig:fig2}, the time series of  $(1-p)N$ active and $pN$ inactive patches are shown for six different $p$ values $p=0,0.15,0.50,0.60,0.70,0.725$. For $p=0.0,$ i.e., all the patches are in active states, the time series of the globally synchronized ($1$ cluster) prey and predator populations are depicted in  Fig.~\ref{fig:fig2}(a). Now, if we set the inactivation ratio $p=0.15$, i.e., $15\%$ of the total patches are in inactive mode, then the amplitude of the prey and predator oscillations is increased and the $2$-cluster solution occurs in active and inactive patches in Fig.~\ref{fig:fig2}(b). Thus the oscillation can be observed in the inactive patches as well. Therefore the dispersal can save the populations from local extinctions. Similarly for two coupled patches with one in active mode and another in inactive mode, the survivability of the inactive patch under dispersal is observed. The results are discussed extensively in the Appendix. If we increase the number of inactive patches by $p=0.5,$ the behavior of active and inactive patches is shown in Fig.~\ref{fig:fig2}(c). Here the time series of the prey and predator populations in both patches are in the chaotic state (which is verified by the maximum Lyapunov exponent later). An interesting phenomenon is observed for a further increased value at $p=0.6$: the prey and predator populations in active patches are in a periodic state with period-2 orbit whereas in inactive patches the behavior is in period 1 [Fig.~\ref{fig:fig2}(d)]. But if we consider $70\%$ of patches of the whole network as the inactive state, the prey and predator populations in active and inactive patches are in period-4 and period-2 states, respectively  [Fig.~\ref{fig:fig2}(e)]. With further increase in the value of $p$ to $0.725,$ the whole network collapses and thus the dispersal cannot save the metapopulation from extinction anymore. The time series of the extinct state is shown in Fig.~\ref{fig:fig2}(f).    

\par   As can be seen from Fig.~\ref{fig:fig2}, inactive patches manage to sustain oscillatory behavior because of the diffusive interactions with all the active patches as long as $p<0.725$. So due to species dispersal, revival of oscillation takes place in extinct patches before a certain critical threshold of $p$, albeit the types and amplitudes of oscillations for different values of $p$ are not the same. Actually, the patchy environment provided more areas for the prey to seek temporary protection. When the prey become extinct locally at one patch, they were able to reestablish themselves by migrating to new patches before being attacked by predators. This habitat spatial structure of patches allowed for coexistence between the predator and prey species and promoted a stable population oscillation model.
\begin{figure}[ht]
	\begin{center}
		\includegraphics[scale=0.48]{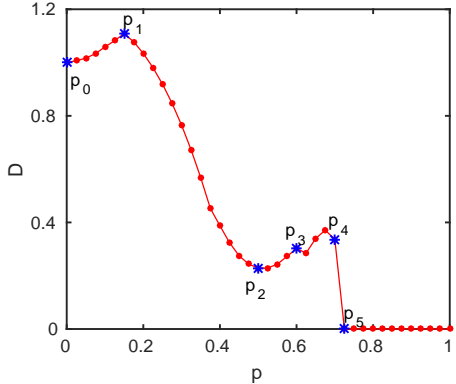}
		\caption{The order parameter $D$ is plotted against the inactivation ratio $p$ for dispersal rate $m=0.07$ with $\alpha= \beta=1.0$. Time series corresponding to the points $p_0, p_1, p_2, p_3, p_4$, and $p_5$ are shown in Fig.~\ref{fig:fig2}, panels (a), (b), (c), (d), (e), and (f), respectively.
			\label{fig:fig3}}
	\end{center}
\end{figure}
\par  To study the macroscopic oscillation level of the entire network, we plot the order parameter $D=\frac{R(p)}{R(0)}$ against the inactivation ratio $p$ in Fig.~\ref{fig:fig3}. From this figure, we can observe that after an initial advancement to the value corresponding to label $p_1$, $D$ decreases up to when $p=0.50$ labeled as $p_2$. After that, $D$ again increases marginally till $p=0.70$ with label $p_4$ and finally drops to zero at a critical value $p_c=0.725$ where the entire network stops oscillating. The meager rise and falls in the values of $D$ have already been described by showing the time series above in Fig.~\ref{fig:fig2} corresponding to all the labeled values of $p$. This basically means that if the inactivation ratio $p$ crosses a critical stage, species dispersal would not be able to revive the population in the inactive patches and the metapopulation goes for extinction. In this case the whole network system  collapses resulting in  extinction of both the prey and predator populations in all the patches. 
\par
\begin{figure}[ht]
	\begin{center}
		\includegraphics[scale=0.45]{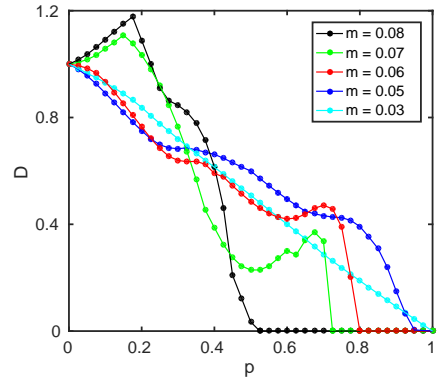}
		\caption{The order parameter $D$ versus the inactivation ratio $p$ in the globally coupled network for different dispersal values of $m$ with $\alpha=\beta=1.0$. The critical ratio $p_c$ at which $D$ becomes zero increases with descending values of the dispersal rate.
			\label{fig:fig4}}
	\end{center}
\end{figure}

 Next, we study the variation of the critical value of the inactivation ratio $p_c$ by varying the dispersal rate $m$.  In Fig.~\ref{fig:fig4}, order parameter $D$ is plotted with respect to the inactivation ratio $p$ for various dispersal rates $m$ taking $\alpha=\beta =1.0$; i.e., prey and predator populations both are having the same dispersal rates. For instance, $m=0.08,0.07,0.06,0.05,0.03$ are considered for figuring out the variations in $D$. Our numerical results show that for $m=0.08$, the critical inactivation ratio for which $D$ turns into zero is $p_c=0.525$. But as $m$ is lowered to $m=0.07$, $p_c$ becomes $0.725$ while $m=0.06, 0.05$ lead to $p_c=0.8, 0.975$, respectively. Finally change in $D$ is shown taking $m=0.03$ that gives $p_c \approx 1$. Thus for decreasing $m$, $p_c$ increases until it reaches unity at a threshold value of $m=0.03$, below which $p_c$ remains at unity. This proves that if the dispersal rate is small enough then even if in $99\%$ of the patches species become extinct the remaining $1\%$ of patches are enough to sustain oscillation in the whole network through dispersal.  Due to small dispersal of species, the active patches act like \textit{refugia} \cite{refugia} that protect the entire species from disturbance. Because of these \textit{refugia}, a fraction of the population survives and small dispersal promotes population persistence over large spatial scales. Currently, some examples of refuge species are the mountain gorilla (isolated to specific mountains in central Africa) and the Australian sea lion. On the other hand, the high dispersal may increase the chance of global extinction in network dynamics. This may be due to the fact that high dispersal leads to fast synchrony of the populations in different patches, which is against the species persistence.  The species-area relationship and dispersal between distinct patches are most important laws in ecology. They can be used to predict species loss under disturbances and are a central tool in conservation biology. Level of clustering has a major role in species survivability. High dispersal leads to a more compact landscape, which is more sensitive to the effect of habitat loss \cite{plosone13}.
 
\par  As illustrated through the basin of attraction of a single-patch model [cf. Fig.~\ref{fig:fig1}(a)], the activity of the patches in terms of oscillatory behavior depends on the choice of initial conditions, for the particular model we have chosen. Because of this,  analytical treatment of the work is cumbersome throughout, but still we elucidate our observation while following a useful approach. Through this, based upon the assumption of $x_i = A_x,~y_i = A_y$ for the active patches and $x_i = I_x,~y_i = I_y$ for the inactive group of patches (basically, synchronized activity among the patches allows one to reformulate the system in such a way), Eq. (\ref{eq:2}) with Eq. (\ref{eq:1}) reduces to the following coupled system: 
	\begin{equation}
	\begin{array}{lcl} \label{eq22}
	\dot{A_x} = \frac{1}{\epsilon}[A_x(1-A_x)(A_x-\theta)-A_xA_y]+ \alpha mp(I_x-A_x),\\
	\dot{I_x} = \frac{1}{\epsilon}[I_x(1-I_x)(I_x-\theta)-I_xI_y]+ \alpha m(1-p)(A_x-I_x),\\
	\dot{A_y} = A_xA_y-d_1A_y+ \beta mp(I_y-A_y),\\
	\dot{I_y} = I_xI_y-d_2I_y+ \beta m(1-p)(A_y-I_y).
	\end{array}
	\end{equation}

	\begin{figure}[ht]
		\begin{center}
			\includegraphics[scale=0.50]{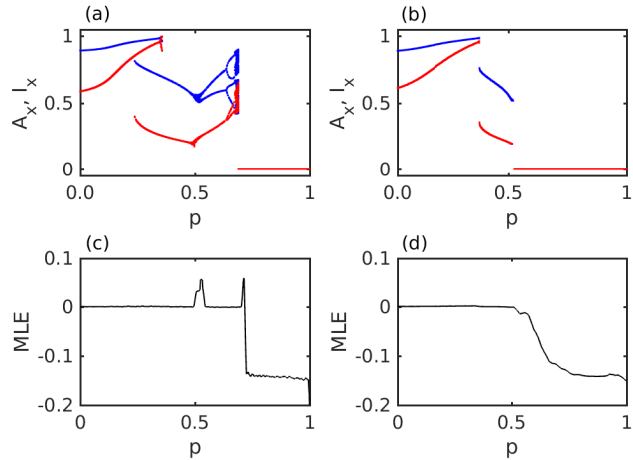}
			\caption{ Bifurcation diagrams of the variables $A_x$ (in blue) and $I_x$ (in red) with respect to the inactivation ratio $p$ for (a) $m=0.07$ and (b) $m=0.08$. The corresponding MLEs against $p$ are plotted in (c) and (d), respectively. The other parameters are the same as in Fig.~\ref{fig:fig2}.
					\label{fig:fig22}}
		\end{center}
	\end{figure}
	
	\par Figure \ref{fig:fig22}(a) shows the bifurcation diagrams of the variables  $A_x,~I_x$ that respectively correspond to the active and inactive prey patches in the reduced model (\ref{eq22}), where the inactivation ratio $p$ has been considered as the bifurcation parameter with $m=0.07$. Here we would like to note that while discussing Fig. \ref{fig:fig2}, we went through qualitative analysis of the time series evolutions of prey and predator patches and the present bifurcation diagrams confirm all the previous observations. For instance, whenever $p=0.5$, the populations exhibit chaotic dynamics whereas for $p=0.6$ and $p=0.7$ active (inactive) patches respectively possess period-2 (period-1) and period-4 (period-2) oscillations, which is quite clear from this bifurcation diagram. Finally, the bifurcation diagram confirms that for $p\ge p_c=0.725$, all the patches go for global extinction. This scenario is further validated by plotting the corresponding maximum Lyapunov exponent (MLE) in Fig. \ref{fig:fig22}(c), where the MLE becomes negative for $p\ge p_c=0.725$. In addition to this, we have also shown the bifurcation diagrams and the associated MLE for a different dispersal strength $m=0.08$ in Figs.\ref{fig:fig22}(b) and (d), respectively. As explained in Fig. \ref{fig:fig4}, with $m=0.08$ the metapopulation goes for extinction at $p_c=0.525$ and here again the bifurcation and MLE diagrams ensure this circumstance.

\begin{figure}[ht]
	\begin{center}
		\includegraphics[scale=0.45]{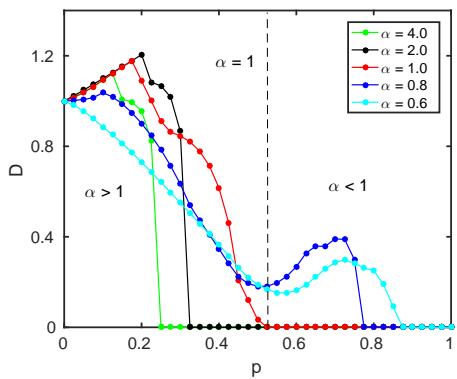}
		\caption{The order parameter $D$ is plotted against $p$ for different $\alpha$ values in a global network with fixed $m = 0.08$ and $\beta=1$. The black dotted line corresponds to the $p_c$ value for $\alpha = 1$. For gradually decreasing value of $\alpha(<1)$, $p_c$ tends towards $1$ and for gradually increasing value of $\alpha(>1)$, $p_c$ tends towards $0$.
			\label{fig:fig5}}
	\end{center}
\end{figure}

\par Now, in Fig. \ref{fig:fig5}, we consider a situation when there is a mismatch in the dispersal rates of prey and predator patches by taking $\beta \not =\alpha$. First we consider $\alpha<\beta$ with $\beta=1$ (which implies that the dispersal rate for prey is much lower than that of the predators) while keeping fixed the value of the dispersal rate $m=0.08$. For $\alpha=0.8,$ that is, the dispersal rate of the predator is larger than the dispersal rate of the prey, the critical value of the inactivation ratio is $p_c=0.775$. This critical value is larger than that for the identical dispersal rate, i.e., $\alpha=\beta=1.0$ (red dotted line). Again with a lower value of $\alpha$ at $0.6,$ the critical value $p_c=0.875$ (cyan dotted line); i.e., enhancement of the survivability of the  metapopulation occurs.  So if the dispersal rate of the  prey is decreased compared to the dispersal rate of the  predator, survivability of the metapopulation increases.   On the contrary, whenever $\alpha>\beta=1$, i.e., the dispersal rate of the prey population is higher than that of the predator, the $p_c$ value gets reduced quite remarkably implying significant de-enhancement in the survivability of the metapopulation. With $\alpha=2.0$, $p_c=0.325$ which was $0.525$ earlier for $\alpha=1$ and an even higher value of  $\alpha=4.0$ drives the metapopulation to achieve $p_c=0.25$. 

In fact, for a particular value of $m$, if we take $\alpha<\beta$, $\beta=1$ or $\beta<\alpha$, $\alpha=1$, i.e., if any one of the asymmetry constants($\alpha$ or $\beta$) is 1 and the other($\beta$ or $\alpha$) is less than 1, then the $p_c$ value increases which effectively makes the metapopulation more persistent. Therefore, mismatch in the dispersal rates of prey and predator helps in species persistence, as long as the values are less ($\alpha, \beta \leq 1$). If we increase the mismatch by taking $\alpha$ (or $\beta$) greater than $1$, keeping the other equal to $1$, then due to fast synchronization the robustness decreases. Also here we mention that as long as $\alpha<1$, the change in the order parameter $D$ is continuous whereas a sort of discontinuous variation is realized in cases of $\alpha>1$ particularly for this model.

\par
\begin{figure}[ht]
	\begin{center}
		\includegraphics[scale=0.45]{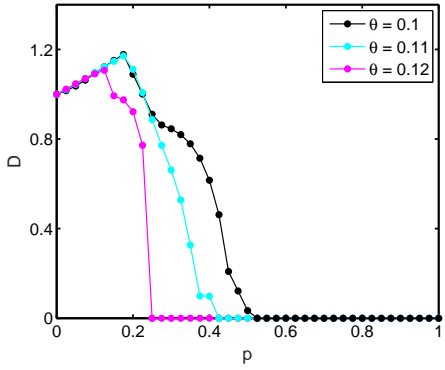}
		\caption{Variation of order parameter $D$ by changing the inactivation ratio $p$ for three different  values of $\theta$  showing de-enhancement of the $p_c$ value for small increment in the $\theta$ value. Other parameters are $m = 0.08$ and $\alpha=\beta=1.0.$
			\label{fig:fig6}}
	\end{center}
\end{figure}

 Next, we will consider another significant parameter in the model, the Allee threshold $\theta$. This Allee threshold is very important because it increases the chance of extinction and a minimum population size is needed for species persistence. The effect of the Allee threshold $\theta$ on the persistence of the metapopulation is shown in Fig.~\ref{fig:fig6}. 
Our study reveals that increasing $\theta$ has negative effects on the order parameter $D$. As discussed above, with $\theta=0.10$, the order parameter $D$ drops to zero at $p_c=0.525$ (in Fig.~\ref{fig:fig4}). However, as we increase the value of $\theta$ to $\theta=0.11$, $D$ declines to zero at $p_c=0.425$ and for further increment in $\theta$, $p_c$ deteriorates to $p_c=0.25$. This basically implies that a strong Allee effect hinders the metapopulation  persistence.
\begin{figure}[ht]
	\begin{center}
		\includegraphics[scale=0.5]{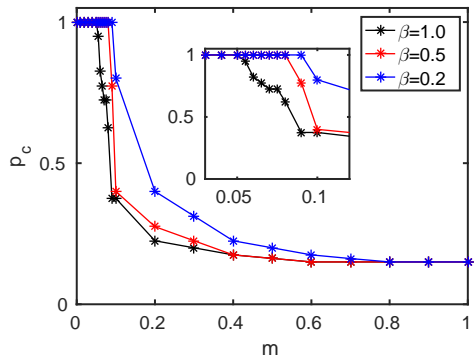}
		\caption{Variation of critical inactivation ratio $p_c$ by changing the dispersal rate $m$ for three different values of $\beta$ in a global network showing enhancement of the $p_c$ value as $\beta$ gradually decreases.
			\label{fig:fig7}}
	\end{center}
\end{figure}
\par In Fig.~\ref{fig:fig7}, we have shown how the critical value $p_c$ varies with dispersal rate $m$ by increasing the dispersal mismatch between prey and predator populations. Keeping $\alpha=1$ fixed, $p_c$ against $m$ is plotted for three different values of $\beta=1.0,0.5$, and $0.2$. As stated earlier, smaller $\beta$ with $\alpha=1$ can be utterly efficient in enhancing metapopulation robustness; here this fact has been shown to be authentic for almost all the values of $m$.  

\begin{figure}[ht]
	\begin{center}
		\includegraphics[scale=0.5]{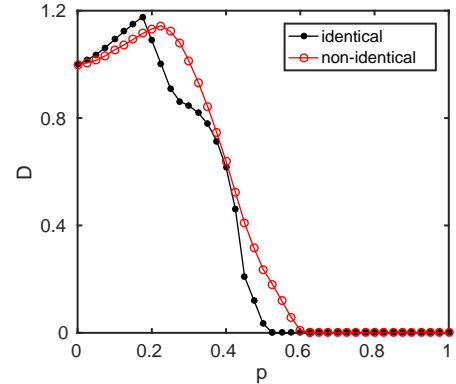}
		\caption{Variation of the order parameter $D$ with respect to inactivation ratio $p$  in a globally coupled network of identical (black) and non-identical (red) patches. Here $m = 0.08$ and $\alpha=\beta=1.0.$
			\label{fig:fig8}}
	\end{center}
\end{figure}

Further we explore the influence of uniformly distributed mortality rate $d$ of the predators in both active and inactive patches, instead of keeping them fixed in those patches. As seen from Fig.~\ref{fig:fig8}, whenever the values of $d$ are taken uniformly from the intervals [0.1, 0.2] and [0.45, 0.54] for inactive and active patches respectively, $p_c$ increases to $0.625$, which was earlier $0.525$ for the fixed identical value of $d$. Due to the parameter mismatch, the population remains in an asynchronous state and metapopulation survivability increases \cite{gupta2017}.

\subsubsection{Complex Dispersal Networks}
\par Next we consider complex dispersal topologies (namely, small world \cite{Watts-Strogatz} and scale free \cite{barabasi199}) among patches and investigate the revival of the population in inactive patches through dispersal. In this context, it is noted that 
the dispersal structures can have substantial influence as far as the metapopulation's persistence is concerned \cite{hanski1999, holland2008,hanski2000}. First we consider a small-world dispersal network. Indeed, earlier studies on plant and animal dispersal structures attest to the fact that most migrating individuals move to shorter distances and some of them travel over longer distances \cite{Ranta2008, dingle2014, turchin1998}. This implies that the majority of the dispersing individuals move to neighboring patches, but there is a probability $q$ that some may arrive at any patch within the network selected at random. Thus in our small-world dispersal, we hold that the species dispersal is not distance dependent between the patches. This fact readily indicates the immense significance of the study of metapopulation survivability under small-world dispersal of the species. We investigate the ecological robustness of patches on small-world dispersal architecture where a fraction of patches initially becomes extinct.
\begin{figure}[ht]
	\begin{center}
		\includegraphics[scale=0.55]{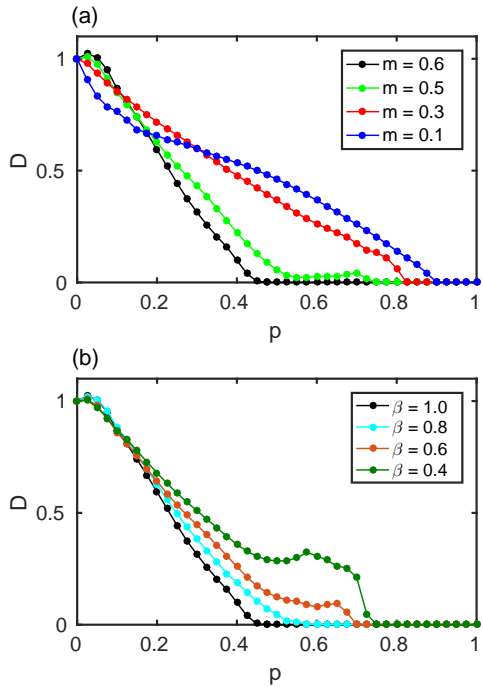}
		\caption{The order parameter $D$ is plotted against the inactivation ratio $p$ for a small-world
			network with average degree $<k>=30$, rewiring probability $q=0.05$: (a) various dispersal rates $m$ with $\alpha=\beta=1.0$,
			and	(b) the effect of the asymmetry parameter $\beta$ where $m = 0.6$ and $\alpha =1.0$.
			\label{fig:fig9}}
	\end{center}
\end{figure}
\par For the small-world model, we consider a network of $N=200$ patches and dispersal matrix $\mathbf{M}$  = $(\frac{\alpha m}{deg(i)}, \frac{\beta m}{deg(i)})^T$, as before. First we consider symmetric dispersal rates between prey and predator populations. In Fig.~\ref{fig:fig9}, the order parameter $D$ is plotted against the inactivation ratio $p$. To find the order parameter $D$ for each $p$, we average over twenty different realizations (even though the results are more or less similar for individual realizations, owing to the randomness in the network architecture, we go for averaging over some realizations).  Like with the global network here also we observe that if the inactivation ratio $p$ crosses a critical stage $p_c$, species dispersal would not be able to resurrect the population in the inactive patches and it goes for extinction in all the patches. Figure~\ref{fig:fig9}(a) shows the variation of $D$ with respect to $p$ for different values of the dispersal rate $m=0.6, 0.5, 0.3$, and $0.1$. As long as the dispersal rate is comparatively high and fixed at $m=0.6$, after a slight initial accession, $D$ starts decreasing monotonically with increasing $p$ and eventually drops to zero for $p_c=0.475$ indicating global extinction of the metapopulation. A similar changeover in $D$ has been witnessed whenever $m=0.5$ is considered, except the fact that the diminution in the value of $D$ is not that hefty this time within $0.525\leq p \leq 0.725$. Nevertheless, the transition possesses a significant difference in the value of $p_c$, which is $0.75$ now. For a lower $m=0.3$, the previously discussed initial hike in the value of $D$ is not there any more and it starts decreasing right from the beginning that becomes zero at $p_c=0.825$. Finally we take $m=0.1$ for which the critical inactivation ratio $p_c=0.90$ which implies that the metapopulation survivability is higher than the previous cases. However, we also note that within the initiatory range of $p$ satisfying $p\lesssim 0.15$, the value of $D$ is lower than that obtained in the earlier cases. In fact, lower dispersal rates allow metapopulation survivability for higher inactivation ratio $p$, but for lower $p$, metapopulation abundance is lower compared to higher dispersal rates.  
\par Next we move on to examine the influence of the parameter $\beta$ that basically controls the asymmetry between the dispersal rates of prey and predators. In order to do this, we figure out the change in $D$ against increasing $p$ for different values of $\beta$ using the fixed value of dispersal rate $m=0.6$ in Fig.~\ref{fig:fig9}(b). As described above, for $\beta=1.0$ (i.e., when the dispersal rates of both species are the  same), the critical inactivation ratio is $p_c=0.475$, but with a decrement in $\beta$ to $\beta=0.8$, the value of $D$ remains nonzero even for all $p<0.60$. This immediately implies that this smaller $\beta$ enhances the survivability of the metapopulation and hence allows more extinct patches to contribute in the metapopulation. For an even smaller $\beta=0.6$, the critical value of the inactivation ratio $p_c$ increases further to $0.7$ indicating more metapopulation persistence. Finally, $p_c$ develops to $0.75$ with $\beta=0.4$. Thus, interestingly enough, the asymmetric dispersal rate achieved in terms of smaller $\beta$ has been investigated to enhance the metapopulation survivability quite comprehensively even for small-world dispersal topology.
\begin{figure}[ht]
	\begin{center}
		\includegraphics[scale=0.55]{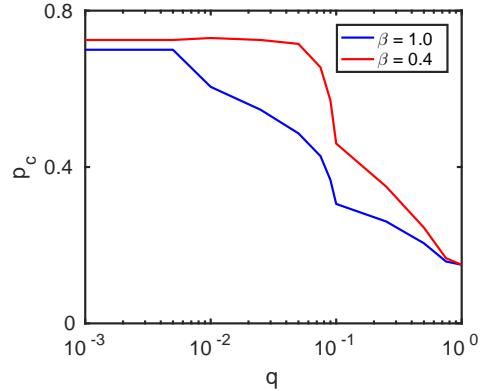}
		\caption{Variation of the $p_c$ values with respect to small-world rewiring probability $q$ for $m = 0.6$ showing deterioration of the critical fraction $p_c$ with increasing randomness in the network.
			\label{fig:fig10}}
	\end{center}
\end{figure}

\par Figure~\ref{fig:fig10} shows how this critical value $p_c$ of the inactivation ratio is correlated with the rewiring probability $q$. When $q=0$, the dispersal network is fully regular; for $0.001<q <0.1$, it is a small-world network possessing low diameter and high clustering. As we increase the value of $q$ further it becomes a more and more random network and at $q=1$ it becomes fully random. For $\alpha=\beta=1.0$, our study affirms that as $q$ increases (implying the presence of many more long range dispersals), initially $p_c$ remains the same for a while but as $q$ crosses a certain value $0.005$, the value of $p_c$ starts decreasing significantly and reaches $0.15$ finally for fully random dispersals. As we consider a lower value of $\beta=0.4$, more or less a similar qualitative scenario is realized, but with higher $p_c$ values throughout. In this case, $p_c$ remains almost the same unless $q>0.05$ and starts decreasing afterwards. This figure firmly points out the efficiency of the asymmetry parameter $\beta$ in augmenting the metapopulation survivability regardless of the rewiring probability (associated with long range dispersals) in the network.  
\begin{figure}[ht]
	\begin{center}
		\includegraphics[scale=0.45]{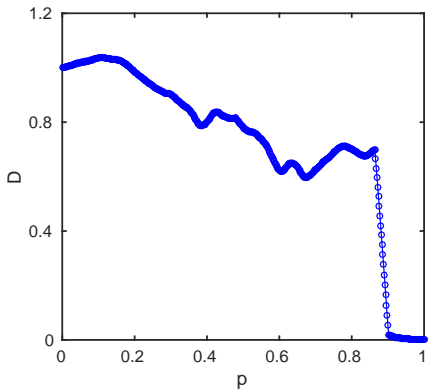}
		\caption{Scale-free network: $D$ is plotted against the inactivation ratio $p$.  Here inactivation of  nodes are  random and  $m=0.1$, $\alpha=\beta=1.0$.
			\label{fig:fig11}}
	\end{center}
\end{figure}

\par Finally, we consider the scale-free dispersal network in which the degree distribution of the network nodes follows the power law \cite{barabasi199} $P(k)\propto k^{-\gamma},$  where $P(k)$ is the probability of finding a node of degree $k$ and $\gamma$ is the power-law exponent (in our case, $\gamma=3$). This basically implies that in most of the patches, species dispersal takes place from a very few patches of the network, but in a few patches dispersal takes place from a large number of patches.  To study the robustness of the prey-predator metapopulation in the scale-free network, we consider three different processes of inactivation in the patches, namely (i) random, (ii) targeted hub (highest degree node), and (iii) targeted low-degree nodes. In the following, we will explore the persistence of the metapopulation in the scale-free network under these three different processes of inactivation. Our main emphasis will be to compare the critical value $p_c$ of the inactivation ratio using these different inactivation processes. 
\begin{figure}[ht]
	\begin{center}
		\includegraphics[scale=0.38]{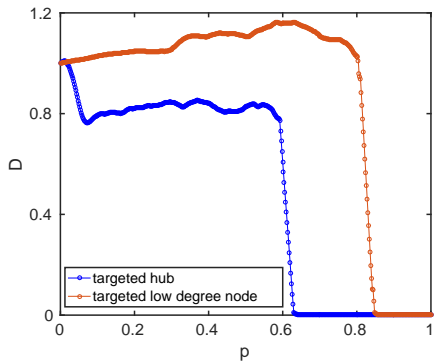}
		\caption{Scale-free network: $D$ is plotted against the inactivation ratio $p$ with targeted high-degree and low-degree nodes sequentially. Here  $m=0.1$, $\alpha=\beta=1.0$.
			\label{fig:fig12}}
	\end{center}
\end{figure}
\par In Fig.~\ref{fig:fig11}, the order parameter $D$ is plotted against the inactivation ratio $p$ when the inactivation process is random.  For this heterogeneous network, the critical value of $p$ for which the  metapopulation goes for extinction remains very high  whenever $m=0.1$. It is also notable that the order parameter $D$ remains very high before it becomes zero. This implies a high abundance of the metapopulation  just before the extinction and sudden extinction of the metapopulation when $90\%$ of the patches are in the  inactive state. 
\par In a heterogeneous network, we have  very few high-degree nodes and many low-degree nodes. To understand the effects of  inactivation of high-degree as well as low-degree nodes  on the order parameter $D$, we make a targeted inactivation of nodes. For this, we inactivate the high (low) degree nodes at first and then inactivate the low (high) degree nodes accordingly. This process is repeated until the number of inactive nodes reach $Np$.  The variation of the order parameter $D$ is plotted against inactivation ratio $p$ for targeted inactivation in Fig.~\ref{fig:fig12}. From this figure one can observe that the critical value $p_c$ is much lower in the case of targeted inactivation of high-degree nodes. It is to be noted that a patch benefits from its nearest neighbor's degree. Thus a patch will benefit more if it is connected to a higher degree patch, and the effect may be larger than the patch's own degree. Therefore, if we target to inactivate the hubs, the metapopulation persistence decreases \cite{Gilarranz2012}. In all the cases, the abundance of the metapopulation is very high until the inactivation ratio $p$ reaches a critical value $p_c$, but in the case of random inactivation the  chances of metapopulation persistence are higher compared to targeted inactivations. 

\section{Holling type II functional response}
As a confirmation of the fact that the obtained results are not model specific, in this section we go through a study of the damaged metapopulation of fit and unfit patches while considering the Rosenzweig-MacArthur prey-predator model possessing the Holling type II functional response as follows: 
\medskip
\begin{equation}\label{eq:4}
\begin{split}
\frac{dx}{dt} & = \frac{1}{\epsilon}\Big [x(1-x)(x-\theta)-\frac{axy}{b+x}\Big]\\
\frac{dy}{dt} & = \frac{axy}{b+x}-dy,
\end{split}
\end{equation}
where $x$ is the prey population density, $y$ is the predator population density,  and $\epsilon$ is the time scale difference between prey and predator population. Here $\theta\in(0,1)$ denotes the Allee threshold, $d$ is the natural mortality rate of the predator population as before, $a$ is the rate of predation, and $b$ is the half-saturation constant. Here the following parameter values have been considered: $\epsilon = 1, \theta = 0.1, a = 1, b = 1$ and $d = 0.28$ (for inactive patches), $d = 0.36$ (for active patches). 

\begin{figure}[ht]
	\centerline{
		\includegraphics[scale=0.5]{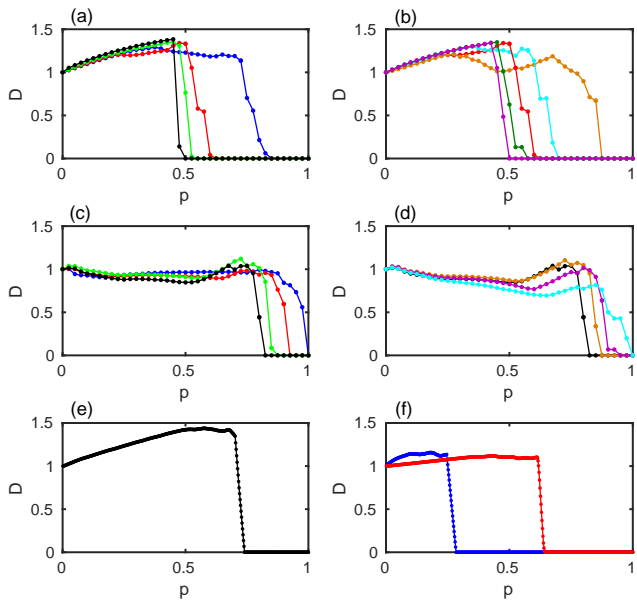}}
	\caption{ Variation of the order parameter $D$ with respect to $p$: globally coupled patches (N=200) (a) for different dispersal rates $m = 0.04$ (blue), $0.05$ (red), $0.06$ (green), $0.07$ (black) with $\alpha, \beta =1$; (b) taking $m$ fixed at $m = 0.05$ and $\beta=1$, the effect of asymmetry parameter $\alpha$ has been shown here, for $\alpha = 1.0$ (red), $0.8$ (cyan), $0.6$ (brown), $1.2$ (green), $1.4$ (violet). Small-world dispersal network (N=200): $p$ vs order parameter $D$ with average degree $<k> = 40$, rewiring probability $q = 0.05$ (c) for different dispersal rates $m = 0.3$ (blue), $0.5$ (red), $0.7$ (green), $0.8$ (black) with $\alpha, \beta =1$; (d) taking $m$ fixed at $m = 0.8$ and $\beta=1$, the effect of asymmetry parameter $\alpha$ has been shown here, for $\alpha = 1.0$ (black), $0.8$ (brown), $0.6$ (violet), $0.4$ (cyan). Scale-free dispersal network (N=500): Order parameter $D$ (e) with random inactivation procedure and (f) targeted inactivation of hubs (blue) and low-degree nodes (red) for $m=0.10$ with $\alpha=\beta=1$. 
		\label{hollingII}}
\end{figure}

\par Figure \ref{hollingII}(a) depicts the variation in the order parameter $D$ with respect to increasing values of the inactivation (non-self-oscillatory) ratio $p$ for different values of the dispersal rate $m$ whenever $\alpha=\beta=1$ in the case of global dispersal topology. With small dispersal strength $m=0.04$, global oscillation in the network disappears for $p\ge p_c=0.85$ and consequently the metapopulation passes through complete extinction. As illustrated above for the Holling type I functional response, here also with higher $m=0.05,~0.06$, and $0.07$, this phase transition in the metapopulation occurs much earlier at $p_c=0.625,~0.55$, and $0.5$, respectively. Next we introduce asymmetry in terms of nonunit $\alpha$ in the dispersal process while keeping $m=0.05$ and $\beta=1$ fixed, as shown in Fig. \ref{hollingII}(b). Due to $\alpha=1.2$, the value of $p_c$ gets lowered to $p_c=0.575$ from $p_c=0.625$ implying faster metapopulation extinction. Higher $\alpha=1.4$ leads to $p_c=0.5$ meaning even when $50\%$ patches are unfit, the metapopulation would eventually collapse. But much more importantly, decreasing $\alpha<1$ may cause significant improvement in the metapopulation survivability. For instance, with $\alpha=0.8$ the metapopulation persists up to $p\le p_c=0.7$ and even lower $\alpha=0.6$ remarkably helps the metapopulation to withstand local patch extinctions unless $p$ reaches $p_c=0.875$. 
\par As far as the complex dispersal topologies are concerned, we analyze both small-world and scale-free dispersal structures one-by-one for the patches described by Eq. (\ref{eq:4}). With an average degree of $<k>=40$ and rewiring probability $q=0.05$, the order parameter $D$ is plotted against $p$ for small-world dispersal in Fig. \ref{hollingII}(c). With dispersal rate $m=0.3$, $D$ vanishes for $p\ge p_c=1.0$ and as can be expected higher $m=0.5,~0.7,~0.8$ cause relatively rapid extinction of the metapopulation at $p_c=0.925,~0.875,~0.825$, respectively, while $\alpha=\beta=1$. Enhancement in the metapopulation resilience is further described by inducing $\alpha<1$ with $m=0.8$ and $\beta=1$. Deviation of $\alpha$ from $1$ to $0.8$ and $0.6$ generates improved $p_c=0.875$ and $p_c=0.95$, respectively. Lowering the value of $\alpha$ to $0.4$, one may have the most robust metapopulation as it exhibits $p_c=1.0$ (cf. Fig. \ref{hollingII}(d)).
\par Finally, we consider a metapopulation model possessing the scale-free dispersal mechanism with a power-law degree (patch dispersal connectivity) distribution having power-law exponent $\gamma=3$. As in the previous case, owing to the heterogeneity in the network structure, here again we follow three different inactivation procedures: (i) random inactivation, (ii) targeted hub, and (iii) targeted low-degree nodes inactivation. Figure \ref{hollingII}(e) describes the change in $D$ for increasing $p$ in which inactivation of the patches is done randomly for $m=0.10$ and $p_c$ becomes $0.74$. But whenever inactivations are done while targeting the hubs and low-degree nodes, $D$ turns into zero and hence the metapopulation collapses at $p_c=0.285$ and $p_c=0.65$, respectively, as in Fig. \ref{hollingII}(f). The random inactivation mechanism is again realized to yield much higher metapopulation survivability compared to targeted attacks.

\section{Discussions and conclusions}
 Characteristically metapopulations are viewed as persistence of sets of populations under the balance between colonization and local extinction \cite{Harrison1991local}. Though the metapopulation theory is developed in many areas of ecological and evolutionary biology, many empirical and theoretical questions remain to be answered about how a metapopulation survives. The extinction of a local population is one of the main characteristics of metapopulation theory \cite{hanski1991single}. In the early stage, some field experiments in plant and insect populations observed local extinctions, which are the basis of current metapopulation theory. From the earlier observations, it was argued that populations may persist through dispersal and recolonization in vacant habitats.
\par
\begin{figure}[ht]
	\begin{center}
		\includegraphics[scale=0.45]{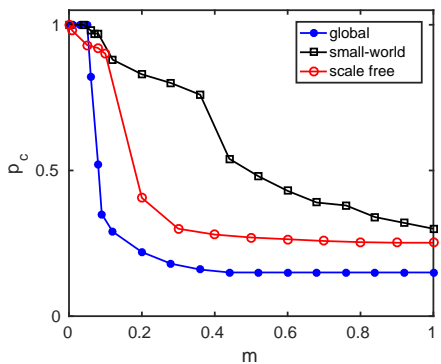}
		\caption{Comparison of the $p_c$ values with respect to the dispersal rate $m$ for three different network topologies, i.e., global, small-world, and scale free. Here the number of patches in the metapopulation is $N=500$ and there is a symmetric dispersal rate, i.e., $\alpha=\beta=1.0$. For small-world network, the average degree $<k>=80$ and rewiring probability $q=0.05$ are taken. Random inactivation procedure has been applied in the case of the scale-free network.
			\label{fig:fig13}}
	\end{center}
\end{figure}

 To study the survivability of a metapopulation under local extinctions, we have considered three different types of dispersal networks, i.e., global, small world and scale free. Figure~\ref{fig:fig13} shows the variation of the critical inactivation ratio $p_c$ with the dispersal rate $m$.  Our investigation reveals that the small-world network is most robust ecologically while in the case of global dispersal, the possibility of metapopulation extinction is much higher than that of the complex dispersal networks. For the small-world dispersal topology, the $p_c$ value decreases gradually with increasing $m$. For the other two cases, the fall in $p_c$ values is quite abrupt; however these values remain almost the same for a long range of $m$ beyond $0.4$ in each case.  From the perspective of persistence of the metapopulation, the small-world dispersal network is more robust compared to global and scale-free dispersals. This is because of the asynchrony maintained among the patches in the case of small-world dispersal  for the whole range of dispersal rate $m$ in $[0, 1]$.   Moreover, our work also shows that in the case of scale-free dispersal, metapopulation abundance is comparatively higher than that of global and small-world dispersals before the complete extinction of the metapopulation. 

\par To conclude, through this article we have scrutinized one of the most important issues in the ecology of survivability of a metapopulation. We have made an attempt to explain up to which level such a metapopulation can persist that passes through several patch extinctions, while considering global, small-world, and scale-free dispersals. In the case of global and small-world topology, we have explored the effect of dispersal rate on the persistence of a metapopulation for both symmetric and asymmetric dispersals. For identical dispersal of prey and predator populations, the ecological patches are more robust for smaller values of the dispersal rate.   When asymmetry is introduced in the dispersal rate, enhancement of ecological robustness is observed if the dispersal rate of the predator (or prey) is lower than that of the prey (or predator) population.  The effect of the Allee threshold on the robustness of a metapopulation is also discussed in the case of global dispersal and it is noted that the survivability increases for smaller values of that threshold. Regarding all-to-all dispersals we have also shown that the metapopulation persistence is higher for nonidentical patches than for identical patches. In the instance of the small-world network we have shown that the critical inactivation ratio decreases gradually with increasing rewiring probability. While considering scale-free dispersal, we have thoroughly investigated how different types of inactivation processes affect the survivability of a metapopulation. Our results state that the survivability is maximum in the case of a random inactivation procedure compared to the targeted inactivation of hubs and low-degree nodes. The survivability is found to be minimum for the targeted inactivation of hubs. To elucidate that the observed scenarios are not model specific, we have performed the investigation of the local dynamical units (prey-predator patches) exhibiting both Holling type I and type II functional responses. This sort of study on dynamical robustness of systems based on naturally possible occurrences is highly relevant as it mimics other realistic scenarios as well. For instance, in the cases of communication in neuronal systems \cite{app1}, proper functioning in cardiac and respiratory systems \cite{app2}, and specific physiological processes \cite{app3} such as cell necrosis within organs \cite{necro22}, stable and robust global oscillation is quite essential. Even from the perspective of power grids \cite{top10} and the El Ni\~{n}o or southern oscillation in Earth's ocean and atmosphere \cite{nino}, the study of dynamical rhythmicity is very important. 
\par Our results indicate a type of behavior in a metapopulation with various dispersal topologies that experiences local extinctions, but the generality of the conclusions is subjected to further studies.  We believe that our results will particularly improve the theoretical understanding of metapopulation persistence of interacting ecological species.  \\

\noindent \textbf{Acknowledgments} \\
D.G. was supported by SERB-DST (Department of Science and Technology), Government of India (Project No. EMR/2016/001039). S.K.S. was supported by NBHM post-doc fellowship.

\section{Appendix}

{\bf Two coupled model: one active and another in inactive mode}
\\
\par We consider two distinct and spatially separated predator-prey patches where both predators and prey disperse between patches with a constant dispersing rate $m$.  Thus our predator-prey model in two patches with diffusive coupling is given by the following equations:
\begin{equation}
\label{eq:3}
\begin{split}
\frac{dx_i}{dt} & =\frac{1}{\epsilon}[x_i(1-x_i)(x_i-\theta)-x_iy_i]+ \alpha m (x_j-x_i)\\
\frac{dy_i}{dt} & =x_iy_i-d_iy_i + \beta m (y_j-y_i)
\end{split}
\end{equation}
where $i,j=1,2$ and $i\neq j$. 
\par  The model equation \eqref{eq:3} without the spatial parameter (i.e., $m=0$) gives the local dynamics of each patch. Here we have taken parameter values in such a manner that one patch (patch 1) gives limit-cycle behavior (active patch, $d_1=0.5$) and the other one (patch 2, $d_2=0.2$) remains in extinction equilibrium (inactive patch). We are interested in seeing whether as a result of species dispersal we can revive the population in patch 2. Before going to numerical simulation, let us discuss the possible equilibrium points of the coupled patches.  The two coupled patches [Eq. \ref{eq:3}] provide five possible equilibrium points and we discuss here the linear stability analysis near these equilibrium points. The existence conditions for the equilibrium points and their local stability are discussed below.

	\begin{figure}[ht]
		\begin{center}
			\includegraphics[scale=0.5]{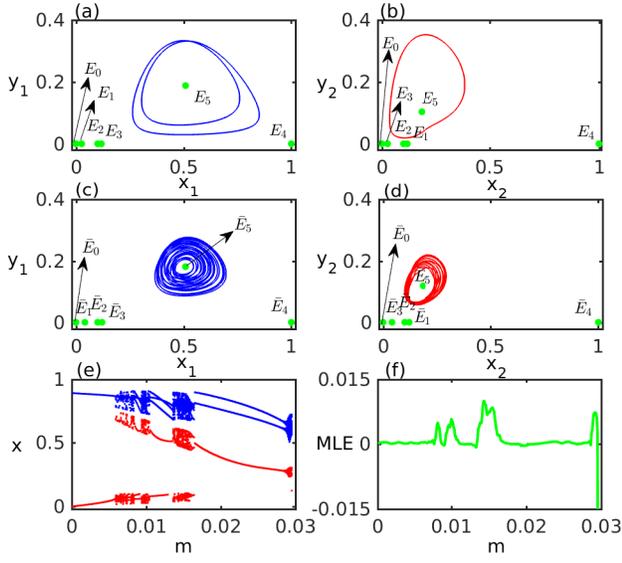}		
			\caption{  Two-coupled model. (a) Period-2, (b) period-1 limit cycle together with the equilibrium points $E_j~(j=0,1,2,3,4,5)$ corresponding to $m=0.02$ respectively for active and inactive patches. Here $E_0=(0,0,0,0)$, $E_1=(0.025,0,0.118,0)$, $E_2=(0.1,0,0.1,0)$, $E_3=(0.118,0,0.025,0)$, $E_4=(1,0,1,0)$, and $E_5=(0.509,0.188,0.183,0.104)$.
					(c), (d) Chaotic orbits along with the equilibrium points $\bar{E}_k~(k=0,1,2,3,4,5)$ corresponding to $m=0.029$ respectively for active and inactive patches. Here $\bar{E}_0=(0,0,0,0)$, $\bar{E}_1=(0.041,0,0.122,0)$, $\bar{E}_2=(0.1,0,0.1,0)$, $\bar{E}_3=(0.122,0,0.041,0)$, $\bar{E}_4=(1,0,1,0)$, and $\bar{E}_5=(0.509,0.182,0.185,0.120)$. (e) Bifurcation diagram by changing the dispersal rate $m$, (f) corresponding maximum Lyapunov exponent with respect to $m$. Blue and red color in (e) corresponds to the active and inactive patch, respectively. Other parameters are $\epsilon=1.0, \theta=0.1, \alpha=\beta=1.0.$}
			\label{fig:fig14}
		\end{center}
	\end{figure}

	\begin{enumerate} 
\item For any parameter values, the model (\ref{eq:3}) always has one trivial equilibrium point:	\begin{equation}
	E_{0000} = 
	\left|
	\begin{array}{l}
	0 \\
	0 \\
	0 \\
	0 
	\end{array}
	\right.
	\end{equation}  
	where all the active and inactive populations go extinct and is always locally asymptotically stable as the eigenvalues corresponding to this equilibrium point are 
	\begin{equation} 
	\Lambda_{0000} = \left|
	\begin{array}{l} 
	-\frac{\theta}{\epsilon}(<0), \\
	-\frac{\theta+2\alpha m \epsilon}{\epsilon}(<0), \\
	\frac{-(d_1+d_2+2\beta m) \pm \sqrt{(d_2-d_1)^2+4\beta^2m^2}}{2}(<0) \\
	\end{array}
	\right.
	\end{equation}
	
\item The equilibrium point 
\begin{equation}
E_{\theta0\theta0} = 
\left|
\begin{array}{l}
\theta \\
0 \\
\theta \\
0 
\end{array}
\right.
\end{equation} 
 where only the prey populations in both active and inactive patches exist and is always unstable as the associated eigenvalues with this equilibrium are
\begin{equation} 
\Lambda_{\theta0\theta0} = \left|
\begin{array}{l} 
\frac{\theta(1-\theta)}{\epsilon}(>0), \\
\frac{\theta(1-\theta)}{\epsilon}-2\alpha m, \\
\end{array}
\right.
\end{equation}
and the other two eigenvalues are the roots of the quadratic equation $\lambda^2+A\lambda+B=0$, where $A= d_1+d_2+2(\beta m -\theta)$ and $B = (\theta-d_1)(\theta-d_2)-\beta m (2\theta - d_1-d_2)$. The roots of this quadratic equation are real negative or complex conjugate with negative real parts if $A>0$ and $B>0$.
\item The equilibrium point 
\begin{equation}
E_{1010} = 
\left|
\begin{array}{l}
1 \\
0 \\
1 \\
0 
\end{array}
\right.
\end{equation}
where prey in both active and inactive patches persist at their highest density. The eigenvalues associated with this equilibrium are
\begin{equation} 
\Lambda_{1010} = \left|
\begin{array}{l} 
-\frac{(1-\theta)}{\epsilon}(<0), \\
-\frac{(1-\theta)}{\epsilon} - 2\alpha m(<0), \\
\end{array}
\right.
\end{equation}
and other two eigenvalues are the roots of the quadratic equation $\lambda^2+A_1\lambda+A_2=0$, where $A_1 = d_1+d_2+2(\beta m-1)$ and $A_2 = (1-d_1)(1-d_2)-\beta m (2-d_1-d_2)$. The equilibrium is locally asymptotically stable if both the roots of the above equation are real negative or complex conjugate with negative real parts, i.e., if $A_1>0$ and $A_2>0$. 
\item The equilibrium point $E_{x0x0}=(x_{10},0,x_{20},0)$, where $x_{20}=\frac{x_{10}(\alpha m \epsilon -(1-x_{10})(x_{10}-\theta))}{\alpha m \epsilon}$ and $x_{10}$ is the positive real root of the equation $(\rho^2-\rho(1+\theta)+(\alpha m \epsilon + \theta))(-\rho^2 + \rho\theta +\alpha m \epsilon)(\rho^2-\rho+\alpha m\epsilon)+\alpha^3 m^3 \epsilon^3 =0$. 
 The eigenvalues corresponding to this equilibrium are given by the roots of the biquadratic equation $(\lambda^2 + B_1 \lambda + B_2)(\lambda^2 + B_3 \lambda + B_4) =0$, where $B_1=\frac{1}{\epsilon}[(2x_{10}-1)(x_{10}-\theta)+x_{10}(x_{10}-1)+(2x_{20}-1)(x_{20}-\theta)+x_{20}(x_{20}-1)+2\alpha m \epsilon]$, $B_2=\frac{1}{\epsilon^2}[(x_{10}-\theta)(1-2x_{10})+x_{10}(1-x_{10})][(x_{20}-\theta)(1-2x_{20})+x_{20}(1-x_{20})]-\frac{\alpha m }{\epsilon}[(x_{10}-\theta)(1-2x_{10})+x_{10}(1-x_{10})+(x_{20}-\theta)(1-2x_{20})+x_{20}(1-x_{20})]$, $B_3=(d_1+d_2-x_{10}-x_{20}+2\beta m)$ and $B_4=(x_{10}-d_1)(x_{20}-d_2)-\beta m (x_{10}+x_{20}-d_1-d_2)$. Therefore, the equilibrium exists if $\alpha m \epsilon > (1-x_{10})(x_{10}-\theta)$ and is locally asymptotically stable if all the roots of the above equation are real negative or complex conjugate with negative real parts, i.e., if $B_1,B_2,B_3,B_4>0$.
 \item The coexistence equilibrium $E_{xyxy}=(x_{10},y_{10},x_{20},y_{20})$, where all four populations survive. As the expression of the coexistence equilibrium point is quite difficult, here we set aside detailed discussion on its stability.  
 
\end{enumerate}

\par Our numerical experiment shows that due to species dispersal revival of oscillation can take place in inactive patches to a great extent. Depending on the dispersal rate this oscillation can be both regular (limit cycle) and irregular (chaotic) as shown in Fig.~\ref{fig:fig14}. Figures \ref{fig:fig14}(a) and \ref{fig:fig14}(b) respectively depict the period-2 and period-1 limit cycle orbits for active and inactive patches along with the equilibrium points $E_j~(j=0,1,2,3,4,5)$ whenever $m=0.02$. On the other hand, the presence of chaotic orbits and the equilibrium points $\bar{E}_k~(k=0,1,2,3,4,5)$ are given in Figs. \ref{fig:fig14}(c) and \ref{fig:fig14}(d) with $m=0.029$ for active and inactive patches, respectively. Finally, Fig.~\ref{fig:fig14}(e) shows the bifurcation diagram with respect to the dispersal rate $m$ with $\alpha=\beta=1.0$ from which periodic and chaotic windows are clearly visible.  For better perception, we have plotted the corresponding maximum Lyapunov exponent against the dispersal rate $m$ in Fig.~\ref{fig:fig14}(f).

%


\begin{thebibliography}{42}%
	\makeatletter
	\providecommand \@ifxundefined [1]{%
		\@ifx{#1\undefined}
	}%
	\providecommand \@ifnum [1]{%
		\ifnum #1\expandafter \@firstoftwo
		\else \expandafter \@secondoftwo
		\fi
	}%
	\providecommand \@ifx [1]{%
		\ifx #1\expandafter \@firstoftwo
		\else \expandafter \@secondoftwo
		\fi
	}%
	\providecommand \natexlab [1]{#1}%
	\providecommand \enquote  [1]{``#1''}%
	\providecommand \bibnamefont  [1]{#1}%
	\providecommand \bibfnamefont [1]{#1}%
	\providecommand \citenamefont [1]{#1}%
	\providecommand \href@noop [0]{\@secondoftwo}%
	\providecommand \href [0]{\begingroup \@sanitize@url \@href}%
	\providecommand \@href[1]{\@@startlink{#1}\@@href}%
	\providecommand \@@href[1]{\endgroup#1\@@endlink}%
	\providecommand \@sanitize@url [0]{\catcode `\\12\catcode `\$12\catcode
		`\&12\catcode `\#12\catcode `\^12\catcode `\_12\catcode `\%12\relax}%
	\providecommand \@@startlink[1]{}%
	\providecommand \@@endlink[0]{}%
	\providecommand \url  [0]{\begingroup\@sanitize@url \@url }%
	\providecommand \@url [1]{\endgroup\@href {#1}{\urlprefix }}%
	\providecommand \urlprefix  [0]{URL }%
	\providecommand \Eprint [0]{\href }%
	\providecommand \doibase [0]{http://dx.doi.org/}%
	\providecommand \selectlanguage [0]{\@gobble}%
	\providecommand \bibinfo  [0]{\@secondoftwo}%
	\providecommand \bibfield  [0]{\@secondoftwo}%
	\providecommand \translation [1]{[#1]}%
	\providecommand \BibitemOpen [0]{}%
	\providecommand \bibitemStop [0]{}%
	\providecommand \bibitemNoStop [0]{.\EOS\space}%
	\providecommand \EOS [0]{\spacefactor3000\relax}%
	\providecommand \BibitemShut  [1]{\csname bibitem#1\endcsname}%
	\let\auto@bib@innerbib\@empty
	\bibitem [{\citenamefont {Pikovsky}\ \emph {et~al.}(2003)\citenamefont
		{Pikovsky}, \citenamefont {Rosenblum},\ and\ \citenamefont
		{Kurths}}]{Kurths-book}%
	\BibitemOpen
	\bibfield  {author} {\bibinfo {author} {\bibfnamefont {A.}~\bibnamefont
			{Pikovsky}}, \bibinfo {author} {\bibfnamefont {M.}~\bibnamefont {Rosenblum}},
		\ and\ \bibinfo {author} {\bibfnamefont {J.}~\bibnamefont {Kurths}},\
	}\href@noop {} {\emph {\bibinfo {title} {Synchronization: a universal concept
				in nonlinear sciences}}},\ Vol.~\bibinfo {volume} {12}\ (\bibinfo
	{publisher} {Cambridge university press},\ \bibinfo {year}
	{2003})\BibitemShut {NoStop}%
	\bibitem [{\citenamefont {Strogatz}(2004)}]{Strogatz}%
	\BibitemOpen
	\bibfield  {author} {\bibinfo {author} {\bibfnamefont {S.}~\bibnamefont
			{Strogatz}},\ }\href@noop {} {\emph {\bibinfo {title} {SYNC: How Order
				Emerges From Chaos In the Universe, Nature, and Daily Life}}}\ (\bibinfo
	{publisher} {Hyperion, New York},\ \bibinfo {year} {2004})\BibitemShut
	{NoStop}%
	\bibitem [{\citenamefont {Tanaka}\ \emph {et~al.}(2014)\citenamefont {Tanaka},
		\citenamefont {Morino}, \citenamefont {Daido},\ and\ \citenamefont
		{Aihara}}]{Tanaka2014}%
	\BibitemOpen
	\bibfield  {author} {\bibinfo {author} {\bibfnamefont {G.}~\bibnamefont
			{Tanaka}}, \bibinfo {author} {\bibfnamefont {K.}~\bibnamefont {Morino}},
		\bibinfo {author} {\bibfnamefont {H.}~\bibnamefont {Daido}}, \ and\ \bibinfo
		{author} {\bibfnamefont {K.}~\bibnamefont {Aihara}},\ }\href@noop {}
	{\bibfield  {journal} {\bibinfo  {journal} {Phys. Rev. E}\ }\textbf {\bibinfo
			{volume} {89}},\ \bibinfo {pages} {052906} (\bibinfo {year}
		{2014})}\BibitemShut {NoStop}%
	\bibitem [{\citenamefont {Tanaka}\ \emph {et~al.}(2015)\citenamefont {Tanaka},
		\citenamefont {Morino},\ and\ \citenamefont {Aihara}}]{Tanaka2015}%
	\BibitemOpen
	\bibfield  {author} {\bibinfo {author} {\bibfnamefont {G.}~\bibnamefont
			{Tanaka}}, \bibinfo {author} {\bibfnamefont {K.}~\bibnamefont {Morino}}, \
		and\ \bibinfo {author} {\bibfnamefont {K.}~\bibnamefont {Aihara}},\ }\enquote
	{\bibinfo {title} {Dynamical robustness of complex biological networks},}\
	in\ \href@noop {} {\emph {\bibinfo {booktitle} {Mathematical Approaches to
				Biological Systems: Networks, Oscillations, and Collective Motions}}},\
	\bibinfo {editor} {edited by\ \bibinfo {editor} {\bibfnamefont
			{T.}~\bibnamefont {Ohira}}\ and\ \bibinfo {editor} {\bibfnamefont
			{T.}~\bibnamefont {Uzawa}}}\ (\bibinfo  {publisher} {Springer Japan},\
	\bibinfo {address} {Tokyo},\ \bibinfo {year} {2015})\ pp.\ \bibinfo {pages}
	{29--53}\BibitemShut {NoStop}%
	\bibitem [{\citenamefont {Sasai}\ \emph {et~al.}(2015)\citenamefont {Sasai},
		\citenamefont {Morino}, \citenamefont {Tanaka}, \citenamefont {Almendral},\
		and\ \citenamefont {Aihara}}]{Tanaka2015a}%
	\BibitemOpen
	\bibfield  {author} {\bibinfo {author} {\bibfnamefont {T.}~\bibnamefont
			{Sasai}}, \bibinfo {author} {\bibfnamefont {K.}~\bibnamefont {Morino}},
		\bibinfo {author} {\bibfnamefont {G.}~\bibnamefont {Tanaka}}, \bibinfo
		{author} {\bibfnamefont {J.~A.}\ \bibnamefont {Almendral}}, \ and\ \bibinfo
		{author} {\bibfnamefont {K.}~\bibnamefont {Aihara}},\ }\href@noop {}
	{\bibfield  {journal} {\bibinfo  {journal} {PLOS ONE}\ }\textbf {\bibinfo
			{volume} {10}},\ \bibinfo {pages} {1} (\bibinfo {year} {2015})}\BibitemShut
	{NoStop}%
	\bibitem [{\citenamefont {Yuan}\ \emph {et~al.}(2017)\citenamefont {Yuan},
		\citenamefont {Aihara},\ and\ \citenamefont {Tanaka}}]{Tanaka2017}%
	\BibitemOpen
	\bibfield  {author} {\bibinfo {author} {\bibfnamefont {T.}~\bibnamefont
			{Yuan}}, \bibinfo {author} {\bibfnamefont {K.}~\bibnamefont {Aihara}}, \ and\
		\bibinfo {author} {\bibfnamefont {G.}~\bibnamefont {Tanaka}},\ }\href@noop {}
	{\bibfield  {journal} {\bibinfo  {journal} {Phys. Rev. E}\ }\textbf {\bibinfo
			{volume} {95}},\ \bibinfo {pages} {012315} (\bibinfo {year}
		{2017})}\BibitemShut {NoStop}%
	\bibitem [{\citenamefont {Daido}\ and\ \citenamefont
		{Nakanishi}(2004)}]{daido2004}%
	\BibitemOpen
	\bibfield  {author} {\bibinfo {author} {\bibfnamefont {H.}~\bibnamefont
			{Daido}}\ and\ \bibinfo {author} {\bibfnamefont {K.}~\bibnamefont
			{Nakanishi}},\ }\href@noop {} {\bibfield  {journal} {\bibinfo  {journal}
			{Phys. Rev. letts.}\ }\textbf {\bibinfo {volume} {93}},\ \bibinfo {pages}
		{104101} (\bibinfo {year} {2004})}\BibitemShut {NoStop}%
	\bibitem [{\citenamefont {Tanaka}\ \emph {et~al.}(2012)\citenamefont {Tanaka},
		\citenamefont {Morino},\ and\ \citenamefont {Aihara}}]{tanaka2012sr}%
	\BibitemOpen
	\bibfield  {author} {\bibinfo {author} {\bibfnamefont {G.}~\bibnamefont
			{Tanaka}}, \bibinfo {author} {\bibfnamefont {K.}~\bibnamefont {Morino}}, \
		and\ \bibinfo {author} {\bibfnamefont {K.}~\bibnamefont {Aihara}},\
	}\href@noop {} {\bibfield  {journal} {\bibinfo  {journal} {Scientific
				reports}\ }\textbf {\bibinfo {volume} {2}},\ \bibinfo {pages} {232} (\bibinfo
		{year} {2012})}\BibitemShut {NoStop}%
	\bibitem [{\citenamefont {Liu}\ \emph {et~al.}(2016)\citenamefont {Liu},
		\citenamefont {Zou}, \citenamefont {Zhan}, \citenamefont {Duan},\ and\
		\citenamefont {Kurths}}]{liu2016}%
	\BibitemOpen
	\bibfield  {author} {\bibinfo {author} {\bibfnamefont {Y.}~\bibnamefont
			{Liu}}, \bibinfo {author} {\bibfnamefont {W.}~\bibnamefont {Zou}}, \bibinfo
		{author} {\bibfnamefont {M.}~\bibnamefont {Zhan}}, \bibinfo {author}
		{\bibfnamefont {J.}~\bibnamefont {Duan}}, \ and\ \bibinfo {author}
		{\bibfnamefont {J.}~\bibnamefont {Kurths}},\ }\href@noop {} {\bibfield
		{journal} {\bibinfo  {journal} {Europhys. Letts.}\ }\textbf {\bibinfo
			{volume} {114}},\ \bibinfo {pages} {40004} (\bibinfo {year}
		{2016})}\BibitemShut {NoStop}%
	\bibitem [{\citenamefont {Tanaka}\ \emph {et~al.}(2010)\citenamefont {Tanaka},
		\citenamefont {Okada},\ and\ \citenamefont {Aihara}}]{tanaka2010p}%
	\BibitemOpen
	\bibfield  {author} {\bibinfo {author} {\bibfnamefont {G.}~\bibnamefont
			{Tanaka}}, \bibinfo {author} {\bibfnamefont {Y.}~\bibnamefont {Okada}}, \
		and\ \bibinfo {author} {\bibfnamefont {K.}~\bibnamefont {Aihara}},\
	}\href@noop {} {\bibfield  {journal} {\bibinfo  {journal} {Phys. Rev. E}\
		}\textbf {\bibinfo {volume} {82}},\ \bibinfo {pages} {035202} (\bibinfo
		{year} {2010})}\BibitemShut {NoStop}%
	\bibitem [{\citenamefont {Daido}\ and\ \citenamefont
		{Nakanishi}(2007)}]{daido2007}%
	\BibitemOpen
	\bibfield  {author} {\bibinfo {author} {\bibfnamefont {H.}~\bibnamefont
			{Daido}}\ and\ \bibinfo {author} {\bibfnamefont {K.}~\bibnamefont
			{Nakanishi}},\ }\href@noop {} {\bibfield  {journal} {\bibinfo  {journal}
			{Phys. Rev. E}\ }\textbf {\bibinfo {volume} {75}},\ \bibinfo {pages} {056206}
		(\bibinfo {year} {2007})}\BibitemShut {NoStop}%
	\bibitem [{\citenamefont {Levin}(1970)}]{levin1970}%
	\BibitemOpen
	\bibfield  {author} {\bibinfo {author} {\bibfnamefont {R.}~\bibnamefont
			{Levin}},\ }\href@noop {} {\bibfield  {journal} {\bibinfo  {journal} {Some
				mathematical problems in biology. American Mathematical Society, Providence,
				Rhode Island}\ ,\ \bibinfo {pages} {77}} (\bibinfo {year}
		{1970})}\BibitemShut {NoStop}%
	\bibitem [{\citenamefont {Ranta}\ \emph {et~al.}(1995)\citenamefont {Ranta},
		\citenamefont {Kaitala}, \citenamefont {Lindstrom},\ and\ \citenamefont
		{Linden}}]{Ranta95}%
	\BibitemOpen
	\bibfield  {author} {\bibinfo {author} {\bibfnamefont {E.}~\bibnamefont
			{Ranta}}, \bibinfo {author} {\bibfnamefont {V.}~\bibnamefont {Kaitala}},
		\bibinfo {author} {\bibfnamefont {J.}~\bibnamefont {Lindstrom}}, \ and\
		\bibinfo {author} {\bibfnamefont {H.}~\bibnamefont {Linden}},\ }\href@noop {}
	{\bibfield  {journal} {\bibinfo  {journal} {Proceedings: Biological
				Sciences}\ }\textbf {\bibinfo {volume} {262}},\ \bibinfo {pages} {113}
		(\bibinfo {year} {1995})}\BibitemShut {NoStop}%
	\bibitem [{\citenamefont {Goldwyn}\ and\ \citenamefont
		{Hastings}(2011)}]{Goldwyn}%
	\BibitemOpen
	\bibfield  {author} {\bibinfo {author} {\bibfnamefont {E.~E.}\ \bibnamefont
			{Goldwyn}}\ and\ \bibinfo {author} {\bibfnamefont {A.}~\bibnamefont
			{Hastings}},\ }\href@noop {} {\bibfield  {journal} {\bibinfo  {journal}
			{Journal of Theoretical Biology}\ }\textbf {\bibinfo {volume} {289}},\
		\bibinfo {pages} {237 } (\bibinfo {year} {2011})}\BibitemShut {NoStop}%
	\bibitem [{\citenamefont {Watts}\ and\ \citenamefont
		{Strogatz}(1998)}]{Watts-Strogatz}%
	\BibitemOpen
	\bibfield  {author} {\bibinfo {author} {\bibfnamefont {D.~J.}\ \bibnamefont
			{Watts}}\ and\ \bibinfo {author} {\bibfnamefont {S.~H.}\ \bibnamefont
			{Strogatz}},\ }\href@noop {} {\bibfield  {journal} {\bibinfo  {journal}
			{Nature}\ }\textbf {\bibinfo {volume} {393}},\ \bibinfo {pages} {409}
		(\bibinfo {year} {1998})}\BibitemShut {NoStop}%
	\bibitem [{\citenamefont {Ranta}\ \emph {et~al.}(2008)\citenamefont {Ranta},
		\citenamefont {Fowler},\ and\ \citenamefont {Kaitala}}]{Ranta2008}%
	\BibitemOpen
	\bibfield  {author} {\bibinfo {author} {\bibfnamefont {E.}~\bibnamefont
			{Ranta}}, \bibinfo {author} {\bibfnamefont {M.~S.}\ \bibnamefont {Fowler}}, \
		and\ \bibinfo {author} {\bibfnamefont {V.}~\bibnamefont {Kaitala}},\
	}\href@noop {} {\bibfield  {journal} {\bibinfo  {journal} {Proceedings of the
				Royal Society of London B: Biological Sciences}\ }\textbf {\bibinfo {volume}
			{275}},\ \bibinfo {pages} {435} (\bibinfo {year} {2008})}\BibitemShut
	{NoStop}%
	\bibitem [{\citenamefont {Gilarranz}\ and\ \citenamefont
		{Bascompte}(2012)}]{Gilarranz2012}%
	\BibitemOpen
	\bibfield  {author} {\bibinfo {author} {\bibfnamefont {L.~J.}\ \bibnamefont
			{Gilarranz}}\ and\ \bibinfo {author} {\bibfnamefont {J.}~\bibnamefont
			{Bascompte}},\ }\href@noop {} {\bibfield  {journal} {\bibinfo  {journal}
			{Journal of Theoretical Biology}\ }\textbf {\bibinfo {volume} {297}},\
		\bibinfo {pages} {11 } (\bibinfo {year} {2012})}\BibitemShut {NoStop}%
	\bibitem [{\citenamefont {Choudhary}\ and\ \citenamefont
		{Sinha}(2015)}]{sinha201}%
	\BibitemOpen
	\bibfield  {author} {\bibinfo {author} {\bibfnamefont {A.}~\bibnamefont
			{Choudhary}}\ and\ \bibinfo {author} {\bibfnamefont {S.}~\bibnamefont
			{Sinha}},\ }\href@noop {} {\bibfield  {journal} {\bibinfo  {journal} {PloS
				one}\ }\textbf {\bibinfo {volume} {10}},\ \bibinfo {pages} {e0145278}
		(\bibinfo {year} {2015})}\BibitemShut {NoStop}%
	\bibitem [{\citenamefont {Levin}(1974)}]{levin1974}%
	\BibitemOpen
	\bibfield  {author} {\bibinfo {author} {\bibfnamefont {S.~A.}\ \bibnamefont
			{Levin}},\ }\href@noop {} {\bibfield  {journal} {\bibinfo  {journal} {The
				American Naturalist}\ }\textbf {\bibinfo {volume} {108}},\ \bibinfo {pages}
		{207} (\bibinfo {year} {1974})}\BibitemShut {NoStop}%
	\bibitem [{\citenamefont {Hastings}(1993)}]{hastings1993}%
	\BibitemOpen
	\bibfield  {author} {\bibinfo {author} {\bibfnamefont {A.}~\bibnamefont
			{Hastings}},\ }\href@noop {} {\bibfield  {journal} {\bibinfo  {journal}
			{Ecology}\ }\textbf {\bibinfo {volume} {74}},\ \bibinfo {pages} {1362}
		(\bibinfo {year} {1993})}\BibitemShut {NoStop}%
	\bibitem [{\citenamefont {Allee}(1931)}]{Allee}%
	\BibitemOpen
	\bibfield  {author} {\bibinfo {author} {\bibfnamefont {W.~C.}\ \bibnamefont
			{Allee}},\ }\href@noop {} {\emph {\bibinfo {title} {Animal aggregations, a
				study in general sociology}}}\ (\bibinfo  {publisher} {Chicago :The
		University of Chicago Press},\ \bibinfo {year} {1931})\ p.\ \bibinfo {pages}
	{452}\BibitemShut {NoStop}%
	\bibitem [{\citenamefont {Franck~Courchamp}\ and\ \citenamefont
		{Gascoigne}(2008)}]{Franck2008}%
	\BibitemOpen
	\bibfield  {author} {\bibinfo {author} {\bibfnamefont {L.~B.}\ \bibnamefont
			{Franck~Courchamp}}\ and\ \bibinfo {author} {\bibfnamefont {J.}~\bibnamefont
			{Gascoigne}},\ }\href@noop {} {\emph {\bibinfo {title} {Allee Effects in
				Ecology and Conservation}}}\ (\bibinfo {year} {2008})\BibitemShut {NoStop}%
	\bibitem [{\citenamefont {Sasmal}\ and\ \citenamefont
		{Ghosh}(2017)}]{Sasmal2017}%
	\BibitemOpen
	\bibfield  {author} {\bibinfo {author} {\bibfnamefont {S.~K.}\ \bibnamefont
			{Sasmal}}\ and\ \bibinfo {author} {\bibfnamefont {D.}~\bibnamefont {Ghosh}},\
	}\href@noop {} {\bibfield  {journal} {\bibinfo  {journal} {Biosystems}\
		}\textbf {\bibinfo {volume} {151}},\ \bibinfo {pages} {8 } (\bibinfo {year}
		{2017})}\BibitemShut {NoStop}%
	\bibitem [{\citenamefont {Van~Voorn}\ \emph {et~al.}(2007)\citenamefont
		{Van~Voorn}, \citenamefont {Hemerik}, \citenamefont {Boer},\ and\
		\citenamefont {Kooi}}]{van2007h}%
	\BibitemOpen
	\bibfield  {author} {\bibinfo {author} {\bibfnamefont {G.~A.}\ \bibnamefont
			{Van~Voorn}}, \bibinfo {author} {\bibfnamefont {L.}~\bibnamefont {Hemerik}},
		\bibinfo {author} {\bibfnamefont {M.~P.}\ \bibnamefont {Boer}}, \ and\
		\bibinfo {author} {\bibfnamefont {B.~W.}\ \bibnamefont {Kooi}},\ }\href@noop
	{} {\bibfield  {journal} {\bibinfo  {journal} {Mathematical biosciences}\
		}\textbf {\bibinfo {volume} {209}},\ \bibinfo {pages} {451} (\bibinfo {year}
		{2007})}\BibitemShut {NoStop}%
	\bibitem [{sch()}]{scheme}%
	\BibitemOpen
	\href@noop {} {\bibinfo  {journal} {The fifth-order Runge-Kutta-Fehlberg
			integration scheme is used to integrate the systems (1) and (3) with
			integration time step 0.01}\ }\BibitemShut {NoStop}%
	\bibitem [{\citenamefont {Abramson}\ and\ \citenamefont
		{Kenkre}(2002)}]{refugia}%
	\BibitemOpen
	\bibfield  {journal} {  }\bibfield  {author} {\bibinfo {author} {\bibfnamefont
			{G.}~\bibnamefont {Abramson}}\ and\ \bibinfo {author} {\bibfnamefont
			{V.}~\bibnamefont {Kenkre}},\ }\href@noop {} {\bibfield  {journal} {\bibinfo
			{journal} {Physical Review E}\ }\textbf {\bibinfo {volume} {66}},\ \bibinfo
		{pages} {011912} (\bibinfo {year} {2002})}\BibitemShut {NoStop}%
	\bibitem [{\citenamefont {Campos}\ \emph {et~al.}(2013)\citenamefont {Campos},
		\citenamefont {Rosas}, \citenamefont {de~Oliveira},\ and\ \citenamefont
		{Gomes}}]{plosone13}%
	\BibitemOpen
	\bibfield  {author} {\bibinfo {author} {\bibfnamefont {P.~R.}\ \bibnamefont
			{Campos}}, \bibinfo {author} {\bibfnamefont {A.}~\bibnamefont {Rosas}},
		\bibinfo {author} {\bibfnamefont {V.~M.}\ \bibnamefont {de~Oliveira}}, \ and\
		\bibinfo {author} {\bibfnamefont {M.~A.}\ \bibnamefont {Gomes}},\ }\href@noop
	{} {\bibfield  {journal} {\bibinfo  {journal} {PloS one}\ }\textbf {\bibinfo
			{volume} {8}},\ \bibinfo {pages} {e66495} (\bibinfo {year}
		{2013})}\BibitemShut {NoStop}%
	\bibitem [{\citenamefont {Gupta}\ \emph {et~al.}(2017)\citenamefont {Gupta},
		\citenamefont {Banerjee},\ and\ \citenamefont {Dutta}}]{gupta2017}%
	\BibitemOpen
	\bibfield  {author} {\bibinfo {author} {\bibfnamefont {A.}~\bibnamefont
			{Gupta}}, \bibinfo {author} {\bibfnamefont {T.}~\bibnamefont {Banerjee}}, \
		and\ \bibinfo {author} {\bibfnamefont {P.~S.}\ \bibnamefont {Dutta}},\
	}\href@noop {} {\bibfield  {journal} {\bibinfo  {journal} {Phys. Rev. E}\
		}\textbf {\bibinfo {volume} {96}},\ \bibinfo {pages} {042202} (\bibinfo
		{year} {2017})}\BibitemShut {NoStop}%
	\bibitem [{\citenamefont {Barab{\'a}si}\ and\ \citenamefont
		{Albert}(1999)}]{barabasi199}%
	\BibitemOpen
	\bibfield  {author} {\bibinfo {author} {\bibfnamefont {A.-L.}\ \bibnamefont
			{Barab{\'a}si}}\ and\ \bibinfo {author} {\bibfnamefont {R.}~\bibnamefont
			{Albert}},\ }\href@noop {} {\bibfield  {journal} {\bibinfo  {journal}
			{science}\ }\textbf {\bibinfo {volume} {286}},\ \bibinfo {pages} {509}
		(\bibinfo {year} {1999})}\BibitemShut {NoStop}%
	\bibitem [{\citenamefont {Hanski}(1999)}]{hanski1999}%
	\BibitemOpen
	\bibfield  {author} {\bibinfo {author} {\bibfnamefont {I.}~\bibnamefont
			{Hanski}},\ }\href@noop {} {\emph {\bibinfo {title} {Metapopulation
				ecology}}}\ (\bibinfo  {publisher} {Oxford University Press},\ \bibinfo
	{year} {1999})\BibitemShut {NoStop}%
	\bibitem [{\citenamefont {Holland}\ and\ \citenamefont
		{Hastings}(2008)}]{holland2008}%
	\BibitemOpen
	\bibfield  {author} {\bibinfo {author} {\bibfnamefont {M.~D.}\ \bibnamefont
			{Holland}}\ and\ \bibinfo {author} {\bibfnamefont {A.}~\bibnamefont
			{Hastings}},\ }\href@noop {} {\bibfield  {journal} {\bibinfo  {journal}
			{Nature}\ }\textbf {\bibinfo {volume} {456}},\ \bibinfo {pages} {792}
		(\bibinfo {year} {2008})}\BibitemShut {NoStop}%
	\bibitem [{\citenamefont {Hanski}\ and\ \citenamefont
		{Ovaskainen}(2000)}]{hanski2000}%
	\BibitemOpen
	\bibfield  {author} {\bibinfo {author} {\bibfnamefont {I.}~\bibnamefont
			{Hanski}}\ and\ \bibinfo {author} {\bibfnamefont {O.}~\bibnamefont
			{Ovaskainen}},\ }\href@noop {} {\bibfield  {journal} {\bibinfo  {journal}
			{Nature}\ }\textbf {\bibinfo {volume} {404}},\ \bibinfo {pages} {755}
		(\bibinfo {year} {2000})}\BibitemShut {NoStop}%
	\bibitem [{\citenamefont {Dingle}(2014)}]{dingle2014}%
	\BibitemOpen
	\bibfield  {author} {\bibinfo {author} {\bibfnamefont {H.}~\bibnamefont
			{Dingle}},\ }\href@noop {} {\emph {\bibinfo {title} {Migration: the biology
				of life on the move}}}\ (\bibinfo  {publisher} {Oxford University Press,
		USA},\ \bibinfo {year} {2014})\BibitemShut {NoStop}%
	\bibitem [{\citenamefont {Turchin}(1998)}]{turchin1998}%
	\BibitemOpen
	\bibfield  {author} {\bibinfo {author} {\bibfnamefont {P.}~\bibnamefont
			{Turchin}},\ }\href@noop {} {\enquote {\bibinfo {title} {Quantitative
				analysis of movement: measuring and modeling population redistribution in
				plants and animals},}\ } (\bibinfo {year} {1998})\BibitemShut {NoStop}%
	\bibitem [{\citenamefont {Harrison}(1991)}]{Harrison1991local}%
	\BibitemOpen
	\bibfield  {author} {\bibinfo {author} {\bibfnamefont {S.}~\bibnamefont
			{Harrison}},\ }\href@noop {} {\bibfield  {journal} {\bibinfo  {journal}
			{Biological journal of the Linnean Society}\ }\textbf {\bibinfo {volume}
			{42}},\ \bibinfo {pages} {73} (\bibinfo {year} {1991})}\BibitemShut {NoStop}%
	\bibitem [{\citenamefont {Hanski}(1991)}]{hanski1991single}%
	\BibitemOpen
	\bibfield  {author} {\bibinfo {author} {\bibfnamefont {I.}~\bibnamefont
			{Hanski}},\ }\href@noop {} {\bibfield  {journal} {\bibinfo  {journal}
			{Biological Journal of the Linnean Society}\ }\textbf {\bibinfo {volume}
			{42}},\ \bibinfo {pages} {17} (\bibinfo {year} {1991})}\BibitemShut {NoStop}%
	\bibitem [{\citenamefont {Lisman}\ and\ \citenamefont
		{Buzs{\'a}ki}(2008)}]{app1}%
	\BibitemOpen
	\bibfield  {author} {\bibinfo {author} {\bibfnamefont {J.}~\bibnamefont
			{Lisman}}\ and\ \bibinfo {author} {\bibfnamefont {G.}~\bibnamefont
			{Buzs{\'a}ki}},\ }\href@noop {} {\bibfield  {journal} {\bibinfo  {journal}
			{Schizophrenia bulletin}\ }\textbf {\bibinfo {volume} {34}},\ \bibinfo
		{pages} {974} (\bibinfo {year} {2008})}\BibitemShut {NoStop}%
	\bibitem [{\citenamefont {Jalife}\ \emph {et~al.}(1998)\citenamefont {Jalife},
		\citenamefont {Gray}, \citenamefont {Morley},\ and\ \citenamefont
		{Davidenko}}]{app2}%
	\BibitemOpen
	\bibfield  {author} {\bibinfo {author} {\bibfnamefont {J.}~\bibnamefont
			{Jalife}}, \bibinfo {author} {\bibfnamefont {R.~A.}\ \bibnamefont {Gray}},
		\bibinfo {author} {\bibfnamefont {G.~E.}\ \bibnamefont {Morley}}, \ and\
		\bibinfo {author} {\bibfnamefont {J.~M.}\ \bibnamefont {Davidenko}},\
	}\href@noop {} {\bibfield  {journal} {\bibinfo  {journal} {Chaos: An
				Interdisciplinary Journal of Nonlinear Science}\ }\textbf {\bibinfo {volume}
			{8}},\ \bibinfo {pages} {79} (\bibinfo {year} {1998})}\BibitemShut {NoStop}%
	\bibitem [{\citenamefont {Koukkari}\ and\ \citenamefont
		{Sothern}(2007)}]{app3}%
	\BibitemOpen
	\bibfield  {author} {\bibinfo {author} {\bibfnamefont {W.~L.}\ \bibnamefont
			{Koukkari}}\ and\ \bibinfo {author} {\bibfnamefont {R.~B.}\ \bibnamefont
			{Sothern}},\ }\href@noop {} {\emph {\bibinfo {title} {Introducing biological
				rhythms: A primer on the temporal organization of life, with implications for
				health, society, reproduction, and the natural environment}}}\ (\bibinfo
	{publisher} {Springer Science \& Business Media},\ \bibinfo {year}
	{2007})\BibitemShut {NoStop}%
	\bibitem [{\citenamefont {Gurtner}\ \emph {et~al.}(2007)\citenamefont
		{Gurtner}, \citenamefont {Callaghan},\ and\ \citenamefont
		{Longaker}}]{necro22}%
	\BibitemOpen
	\bibfield  {author} {\bibinfo {author} {\bibfnamefont {G.~C.}\ \bibnamefont
			{Gurtner}}, \bibinfo {author} {\bibfnamefont {M.~J.}\ \bibnamefont
			{Callaghan}}, \ and\ \bibinfo {author} {\bibfnamefont {M.~T.}\ \bibnamefont
			{Longaker}},\ }\href@noop {} {\bibfield  {journal} {\bibinfo  {journal}
			{Annu. Rev. Med.}\ }\textbf {\bibinfo {volume} {58}},\ \bibinfo {pages} {299}
		(\bibinfo {year} {2007})}\BibitemShut {NoStop}%
	\bibitem [{\citenamefont {Buldyrev}\ \emph {et~al.}()\citenamefont {Buldyrev},
		\citenamefont {Parshani}, \citenamefont {Paul}, \citenamefont {Stanley},\
		and\ \citenamefont {Havlin}}]{top10}%
	\BibitemOpen
	\bibfield  {author} {\bibinfo {author} {\bibfnamefont {S.~V.}\ \bibnamefont
			{Buldyrev}}, \bibinfo {author} {\bibfnamefont {R.}~\bibnamefont {Parshani}},
		\bibinfo {author} {\bibfnamefont {G.}~\bibnamefont {Paul}}, \bibinfo {author}
		{\bibfnamefont {H.~E.}\ \bibnamefont {Stanley}}, \ and\ \bibinfo {author}
		{\bibfnamefont {S.}~\bibnamefont {Havlin}},\ }\href@noop {} {\bibfield  {journal} {\bibinfo
			{journal} {Nature (London)}\ }\textbf {\bibinfo {volume} {464}},\
		\bibinfo {pages} {1025} (\bibinfo {year} {2010})}\BibitemShut {NoStop}%
	\bibitem [{\citenamefont {Boutle}\ \emph {et~al.}(2007)\citenamefont {Boutle},
		\citenamefont {Taylor},\ and\ \citenamefont {R{\"o}mer}}]{nino}%
	\BibitemOpen
	\bibfield  {journal} {  }\bibfield  {author} {\bibinfo {author} {\bibfnamefont
			{I.}~\bibnamefont {Boutle}}, \bibinfo {author} {\bibfnamefont {R.~H.}\
			\bibnamefont {Taylor}}, \ and\ \bibinfo {author} {\bibfnamefont {R.~A.}\
			\bibnamefont {R{\"o}mer}},\ }\href@noop {} {\bibfield  {journal} {\bibinfo
			{journal} {American Journal of Physics}\ }\textbf {\bibinfo {volume} {75}},\
		\bibinfo {pages} {15} (\bibinfo {year} {2007})}\BibitemShut {NoStop}%
\end{thebibliography}
\end{document}